\documentclass[aps,prb,twocolumn,superscriptaddress,noshowpacs,footinbib]{revtex4-2}

\usepackage{amsmath}
\usepackage[normalem]{ulem}
\usepackage[T1]{fontenc}
\usepackage[hidelinks,colorlinks=true,allcolors=blue,citecolor=blue]{hyperref}
\usepackage{color,graphicx}
\usepackage[table,xcdraw,dvipsnames]{xcolor}
\usepackage{textcomp}
\usepackage{lipsum}
\usepackage{tabularx}
\newcolumntype{C}{>{\centering\arraybackslash}p{1.5cm}}
\newcolumntype{R}{>{\raggedleft\arraybackslash}p{1.5cm}}
\usepackage{multirow}
\usepackage{orcidlink}
\usepackage{amsfonts}

\newcommand{\bono}{3,5-bis($N$-$tert$-butylaminoxyl)-3'-nitro-1,1'-biphenyl}

\newcommand{\blue}[1]{\textcolor{black}{#1}}

\begin{document}
\title{Properties of an organic model $S=1$ Haldane chain system}
	
\author{Ivan Jakovac\,\orcidlink{0009-0005-0393-3601}}
\affiliation{Department of Physics, Faculty of Science, University of Zagreb, Bijenička c. 32, 10000 Zagreb, Croatia}

\author{Tonči Cvitanić}
\affiliation{Department of Physics, Faculty of Science, University of Zagreb, Bijenička c. 32, 10000 Zagreb, Croatia}

\author{Denis Arčon}
\affiliation{Jozef Stefan Institute, Jamova cesta 39, SI-1000 Ljubljana, Slovenia}
\affiliation{Faculty of mathematics and physics, University of Ljubljana, Jadranska 19, SI-1000 Ljubljana, Slovenia}

\author{Mirta Herak\,\orcidlink{0000-0003-3834-8160}}
\affiliation{Institute of Physics, Bijenička c. 46, 10000 Zagreb, Croatia}

\author{Dominik Cin{\v c}ić\,\orcidlink{0000-0002-4081-2420}}
\affiliation{Department of Chemistry, Faculty of Science, University of Zagreb, Horvatovac 102a, 10000 Zagreb, Croatia}

\author{Nea Baus Topić\,\orcidlink{0000-0003-1798-9348}}
\affiliation{Department of Chemistry, Faculty of Science, University of Zagreb, Horvatovac 102a, 10000 Zagreb, Croatia}

\author{Yuko Hosokoshi\,\orcidlink{0000-0002-4480-4137}}
\affiliation{Department of Physics, Osaka Metropolitan University, Osaka, Japan}
\affiliation{Department of Physical Science, Osaka Prefecture University, Osaka, Japan}

\author{Toshio Ono}
\affiliation{Department of Physics, Osaka Metropolitan University, Osaka, Japan}
\affiliation{Department of Physical Science, Osaka Prefecture University, Osaka, Japan}

\author{Ken Iwashita}
\affiliation{Department of Physical Science, Osaka Prefecture University, Osaka, Japan}

\author{Nobuyuki Hayashi}
\affiliation{Department of Physics, Osaka Metropolitan University, Osaka, Japan}

\author{Naoki Amaya}
\affiliation{Department of Physics, Osaka Metropolitan University, Osaka, Japan}

\author{Akira Matsuo\,\orcidlink{0000-0001-5649-9777}}
\affiliation{Institute for Solid State Physics, University of Tokyo, 5-1-5 Kashiwanoha, Kashiwa, Chiba 277-8581, Japan}

\author{Koichi Kindo}
\affiliation{Institute for Solid State Physics, University of Tokyo, 5-1-5 Kashiwanoha, Kashiwa, Chiba 277-8581, Japan}

\author{Ivor Lončarić\,\orcidlink{0000-0002-5554-4641}}
\affiliation{Ru{\dj}er Bošković Institute, Bijenička c. 54, 10000 Zagreb, Croatia}

\author{Mladen Horvati{\'c}\,\orcidlink{0000-0001-7161-0488}}
\affiliation{Laboratoire National des Champs Magn\'{e}tiques Intenses, LNCMI-CNRS (UPR3228), EMFL, Universit\'{e} \\ Grenoble Alpes, UPS and INSA Toulouse, Bo\^{\i}te Postale 166, 38042 Grenoble Cedex 9, France}

\author{Masashi Takigawa\,\orcidlink{0000-0002-4516-5074}}
\affiliation{Institute for Solid State Physics, University of Tokyo, 5-1-5 Kashiwanoha, Kashiwa, Chiba 277-8581, Japan}

\author{Mihael S. Grbić\,\orcidlink{0000-0002-2542-2192}}
\email{corresponding author: mgrbic.phy@pmf.hr}
\affiliation{Department of Physics, Faculty of Science, University of Zagreb, Bijenička c. 32, 10000 Zagreb, Croatia}

\date{\today} 
\begin{abstract}
We present the properties of a new organic $S=1$ antiferromagnetic chain system $m$-NO$_2$PhBNO (abbreviated BoNO). In this biradical system, two unpaired electrons from aminoxyl groups are strongly ferromagnetically coupled ($|J_\text{FM}| \gtrsim 500$~K) which leads to the formation of an effective $S=1$ state for each molecule. The chains of BoNO biradicals propagate along the crystallographic $a$ axis. Temperature dependencies of the $g$ factor and electron paramagnetic resonance (EPR) linewidth are consistent with a low-dimensional system with antiferromagnetic interactions. The EPR data further suggest that BoNO is the first known Haldane system with an almost isotropic $g$ factor ($2.0023 \pm 2$~\textperthousand). The magnetization measurements in magnetic fields up to $40$~T and low-field susceptibility, together with $^1$H nuclear magnetic resonance (NMR) spectra, reveal a dominant intrachain antiferromagnetic exchange coupling of $J_\text{1D}/k_B = (11.3\pm0.1)$~K, and attainable critical magnetic fields of $\mu_0  H_\text{c1} \approx  2$~T and $\mu_0 H_\text{c2} \approx  33$~T. These measurements therefore suggest that BoNO is a unique Haldane system with extremely small magnetic anisotropy. Present results are crucial for a future in-depth NMR study of the low-temperature Tomonaga-Luttinger liquid (TLL) and magnetic field-induced phases, which can be performed in the entire phase space.
\end{abstract}
	
\maketitle

\section{Introduction}
	
In low-dimensional quantum spin systems enhanced quantum fluctuations enable a multitude of ground states that cannot be realized in three dimensional systems. The concurrent development of theoretical analyses and experimental realizations of these systems has enabled great leaps in the understanding of many-body quantum effects. Ground state properties can be modeled by considering the spin value, geometry and strength of interactions, and the interplay of various types of interactions. Testing of such models allows a thorough examination of quantum phenomena; indeed a multitude of knowledge has been gained by studying the ideal $S=1/2$ chain material Cu(C$_4$H$_4$N$_2$)(NO$_3$)$_2$ (CuPzN)~\cite{Kuhne,BreunigScience}; the transverse field Ising chain system CoNb$_2$O$_6$~\cite{ColdeaScience,KinrossPRX}; the strong-rung and strong-leg two-leg spin-ladder systems (C$_5$H$_{12}$N)$_2$CuBr$_4$ (BPCB)~\cite{Patyal,Watson2001,Klanjsek2008} and (C$_7$H$_{10}$N)$_2$CuBr$_4$ (DIMPY)~\cite{Shapiro2007,Jeong2013,Jeong2016}, respectively; and the large-$D$ system NiCl$_2$-$4$SC(NH$_2$)$_2$ (DTN)~\cite{Paduan-Filho1981,Mukhopadhyay2012,Zvyagin2007}. The last three systems were critical in the context of studying magnetic field-induced long-range order of Bose-Einstein condensate (BEC) type~\cite{Zapf2006,Blinder2017,Jeong2017}.\\
\indent Among the chain models, a unique ground state emerges for a simple Hamiltonian $H=J_\text{1D} \Sigma_{ij} \mathbf{S}_i \mathbf{S}_j$ with integer-valued spins $\mathbf{S}_i$, where nearest neighbors interact via Heisenberg antiferromagnetic interaction $J_\text{1D} $. F. D. M. Haldane has shown~\cite{Haldane1983,HaldanePRL} that in this case the ground state is singlet with an energy gap $ \Delta$ to the first excited (triplet) state. This state, also called the Haldane phase, is topologically distinct from its half-integer spin counterpart in which a gapless Tomonaga-Luttinger liquid (TLL) forms. Haldane phase has been shown to be stable in the presence of \blue{a} finite interchain coupling and single-ion anisotropy~\cite{Sakai1990,Wierschem20142}, which are typically unavoidable in any realistic material. Over the years there have been many candidate materials for the realization of the Haldane state, with $S=1$ spin borne by transition-metal ions such as archetypical CsNiCl$_3$, AgVP$_2$S$_6$, Ni(C$_5$D$_{14}$N$_2$)$_2$N$_3$(PF$_6$) (NDMAP), Ni(C$_2$H$_8$N$_2$)NO$_2$(ClO$_4$) (NENP), and PbNi$_2$V$_2$O$_8$. However, although they all host the Haldane phase in zero magnetic field, they are not suitable~\cite{Broholm2002,Ruegg2003} for expanding the research and studying the evolution of the Haldane phase into the magnetic field-induced BEC. To realize the BEC phase, an external magnetic field must first be applied to lower the energy of the triplet state and close the (Haldane) gap at $\mu_0 H_{c1} = \Delta / g\mu_B$, which reestablishes the TLL state. If the spin system is U(1) symmetry invariant, the BEC ground state emerges with further increase of the magnetic field until the system becomes fully polarized at $\blue{g} \mu_0 \blue{\mu_B} H_{c2} \approx 4 \blue{k_B} J_\text{1D} $. The latter symmetry condition is difficult to fulfill since the spin-orbit effects of transition-metal ions add symmetry-breaking terms into the Hamiltonian that cannot be neglected. The available number of candidates is further reduced to none when one looks for those where magnetic interaction $J_\text{1D} $ allows for the entire phase diagram to be accessible in a laboratory.\\
\indent Recently, molecule-based magnetic systems have attracted attention~\cite{Williams2020,TinNatComm} as one can control the energy scale of exchange interactions. Among them, purely organic radical spin systems are of particular interest, as the delocalized nature of electron orbitals ensures the isotropy of the spin Hamiltonian and spin-orbit coupling is weak compared to transition metal systems. The high-spin ($S \geq 1$) systems are formed due to favorable intramolecular spin couplings in certain stable nitroxide-based radicals~\cite{Mukai1975, Kanno1993, Kanno19932, Kumagai1999, Fisher2005}. \blue{Relatively simple $\pi$-conjugated biradicals with favorable intramolecular ferromagnetic coupling function as $S = 1$ building blocks for modeling spin chain systems}~\cite{Rajca1994}. For example, $m$-phenylene-bis($N$-$tert$-butylaminoxyl), abbreviated as BNO, has been shown to have a large ferromagnetic intramolecular interaction of  $|J_\text{FM}|\blue{/k_B} \ge 600$~K~\cite{Hosokoshi2001,Katoh2002}.
When arranged into different crystal lattices, these molecules form an effective one-dimensional (1D) or two-dimensional (2D) system~\cite{Hosokoshi1999, Goddard2012}. Here, the passive substituents are used to define the crystal structure and the apparent dimensionality \blue{of the system}. In contrast to numerous manifestations of the Haldane systems in organometallics,  no $S=1$ nitroxide system has so far displayed suitable quasi-1D geometry~\cite{Fisher2005}.\\
\indent In this study, we present a novel molecular radical system \bono\ ($m$-NO$_2$PhBNO, hereafter abbreviated as BoNO). Our experimental data demonstrates that BoNO is an excellent candidate for a Haldane system, due to BNO $S=1$ units arranged in chains, with intermolecular antiferromagnetic coupling $J_\text{1D}\blue{/k_B} \approx 11.3$~K, which is almost two orders of magnitude smaller from the intramolecular $|J_\text{FM}|$. The molecular and crystal structure is determined by the high-resolution single-crystal X-ray measurements. We have combined multiple experimental techniques: bulk magnetization, \blue{magnetic} susceptibility, electron paramagnetic resonance (EPR) measurements, and density functional theory (DFT) calculations. When the data are compared \blue{to} the existing theoretical density matrix renormalization group (DMRG) and quantum Monte Carlo(QMC) calculations, we demonstrate that BoNO is a fitting candidate for an isotropic Haldane system. Additionally, $^1$H nuclear magnetic resonance (NMR) spectral analysis reveals a strong hyperfine coupling between the $S=1$ system and the probed $^1$H nuclei, paving the way for an extensive $^1$H NMR research of the low-temperature spin dynamics in TLL and BEC phases.

\section{Methods}
BoNO was synthesized by following the conventional procedure~\cite{Kanno1993} starting from 1,3,5-tribromobenzene, via Suzuki coupling with 3-nitrophenylboronic acid, followed by treatment with freshly prepared Ag$_2$O. The obtained material was purified by column chromatography on silica gel with diethyl ether and $n$-hexane. Recrystallization of slow evaporation from a concentrated mixed solution of diethyl ether and $n$-hexane at room temperature yielded red plate single crystals of BoNO. The typical crystal size was $4 \times 1 \times 0.2$~mm$^3$.

The crystallographic data were collected on an Oxford Xcalibur 3 diffractometer using the Mo K$\alpha$ ($\lambda = 0.71$~\AA) radiation source. The $2 \times 1.5 \times 0.2$~mm$^3$ sample was epoxy-glued on a 4-circle $\kappa$-goniometer and measured at $T=100$~K and at $T=300$~K in a dry nitrogen current. The data collection and cell refinement \blue{were} performed by CrysAlis software, the structure was solved by the direct methods and refined anisotropically using SHELX software suite~\cite{Shelx}. The missing hydrogen atom positions were refined using \blue{the} universal force field molecular optimization.\\
\indent DFT calculations were performed using a plane-wave code Quantum ESPRESSO~\cite{Giannozzi2017} with the vdW-DF-cx exchange-correlation functional~\cite{Berland2014} and the so-called GBRV pseudopotentials~\cite{Garrity2014}, with an energy cut-off of 40~Ry for the plane-wave basis set. Atom positions were relaxed until forces on each atom were below 0.03~eV/\AA, with the unit cell  fixed to the experimental one.\\
\indent  The low-field dc magnetization measurements were performed using the Quantum Design MPMS3 magnetometer on $m = 69 \pm 6$~$\mu$g sample. Magnetic susceptibility was obtained by dividing the measured magnetization by the applied magnetic field and multiplying by Mmol/m (Mmol= molar mass, m = mass of the sample). \\
\indent The high-field magnetization at pulsed magnetic fields of up to about 40 T was measured using a nondestructive pulse magnet at the Institute for Solid State Physics at the University of Tokyo. All experiments were performed using small randomly oriented single crystals with typical dimensions of $1 \times 1 \times 0.2$~mm$^3$.\\
\indent X-band ($\sim$9.6~GHz) EPR experiments were performed on cooling of BoNO single crystals mounted on a glass sample holder and sealed under a dynamic vacuum in standard Suprasil quartz tubes ($OD = 4$~mm) with a Bruker Elexsys E500 EPR spectrometer equipped with a TEM104 dual-cavity resonator, and an Oxford Instruments ESR900 cryostat with an ITC503 temperature controller (stability $\pm 0.05$~K). The typical microwave power was set to 0.1~mW while the modulation field was set to the amplitude of 0.02 ~mT and \blue{a} modulation frequency \blue{of} 100~kHz.\\
\indent For the NMR study, we used a rod-shaped BoNO single crystal ($5 \times 0.6 \times 0.3$~mm) placed in a two-axis goniometer probe~\cite{Cvitanic2019} with the longest dimension along the coil axis.  NMR data were collected using an Apollo Tecmag spectrometer, using the $\pi/2 - \tau - \pi$ spin echo sequence, with \blue{an} average $\pi/2$ pulse length \blue{of} $t_{\pi/2}=0.8$~$\mathrm{\mu}$s. 

\section{Results and discussion}
\subsection{Structure and DFT calculations}
\begin{table}[!b]
	\centering	
	\begin{tabularx}{0.8\columnwidth}{lc|c}
		\hline
		& \multicolumn{2}{c}{BoNO} \\ \hline
		Formula unit (f.u.) & \multicolumn{2}{c}{C$_{20}$H$_{25}$N$_3$O$_4$} \\
		Crystal system & \multicolumn{2}{c}{orthorhombic} \\
		Space group & \multicolumn{2}{c}{No.61 $Pbca$} \\
		Point group & \multicolumn{2}{c}{$mmm$} \\
		F.u. in unit cell ($Z$) & \multicolumn{2}{c}{$8$} \\
		Radiation source & \multicolumn{2}{c}{Mo K$_\alpha$}\\
		\hline
		Temperature [K] & $100$ & $300$ \\
		$a$~[\AA] & $10.57900(5)$ & $10.7682(2)$ \\
		$b$~[\AA] & $31.50380(12)$ & $31.6111(8)$ \\
		$c$~[\AA] & $11.68320(5)$ & $11.7298(3)$ \\
		$\rho_\text{calc.}$~[g cm$^{-3}$] & $1.267$ & $1.236$ \\
		Used/total reflections & $3147/4223$ & $3515/4262$\\
		%Parameters refined & $245$ & $250$\\
		$R [I > 2\sigma (I )]$ & $0.0924$ & $0.0556$\\
		Goodness of fit & $1.329$ & $1.051$ \\ \hline
	\end{tabularx}
	\caption{Summary of BoNO's crystallographic data measured at $T=100$ ~K and $T=300$~K in a dry nitrogen current.}
	\label{tab:crystalographic_data}
\end{table}

\begin{figure}[t!]
	\includegraphics[width=0.48\textwidth]{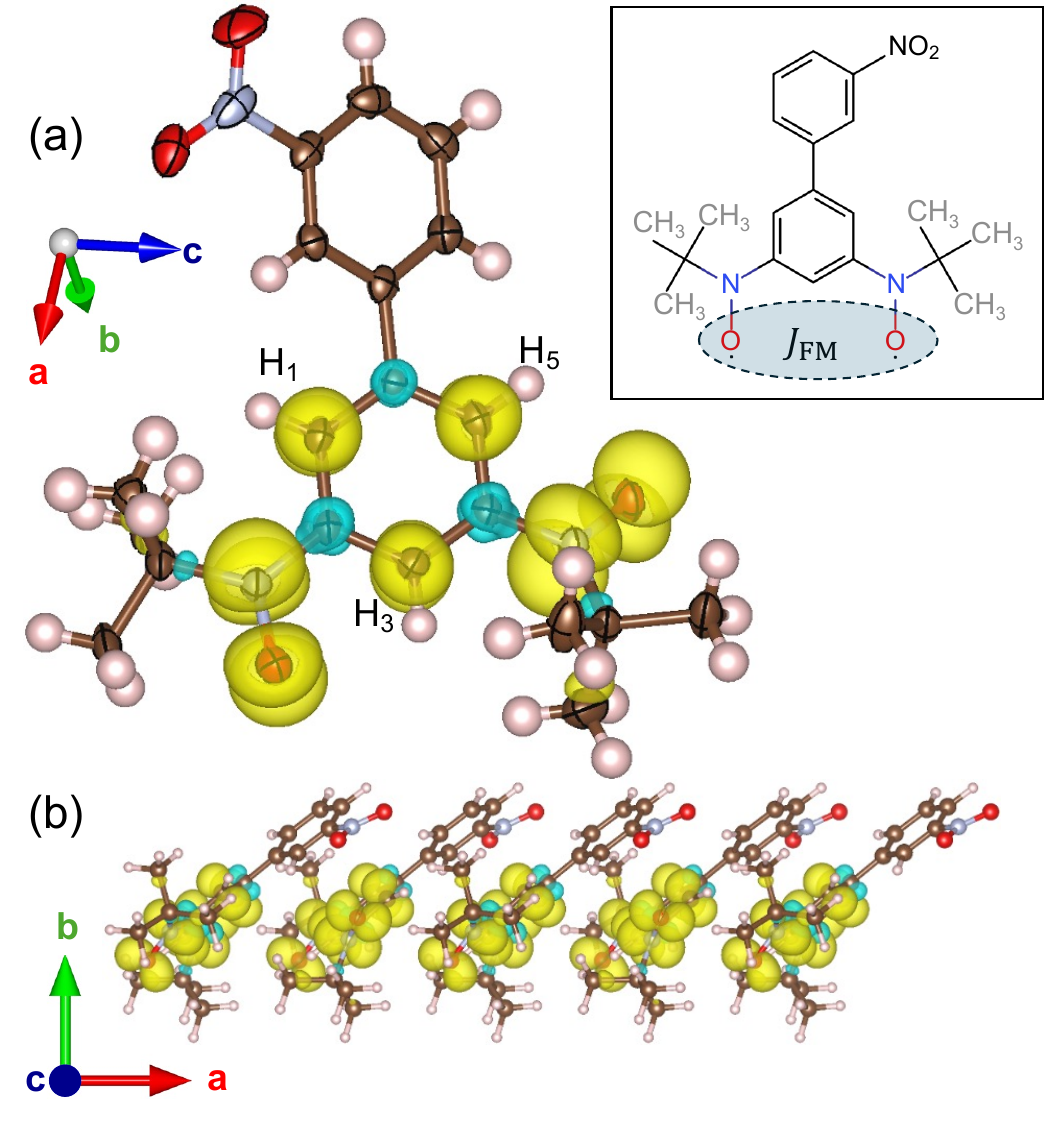}
	\caption{(a) Molecular structure of BoNO - atomic positions are determined by the high-resolution X-ray diffraction with $50 \%$ thermal displacement ellipsoids shown for non-hydrogen atoms: brown (C), pink (H), red (O) and white (N). The positions of hydrogen atoms are determined by universal force field molecular optimization, consistent with non-hydrogen atomic positions isolated from X-ray data ($E_\text{min.} = 463.37$~kJ/mol). The yellow and blue shapes correspond to the positive and negative spin densities, with a cutoff value of  $3\cdot 10^{-3} e$ Bohr$^{-3}$. Crystallographic sites denoted as H$_1$, H$_3$ and H$_5$ posses considerable nuclear hyperfine shifts, making them excellent candidates for an NMR study. \blue{The} inset shows the molecular structure of BoNO; the ferromagnetic coupling unit is emphasized by the shaded dashed oval. (b) \blue{The} stacking of the BoNO molecules to form a chain along the crystallographic \textit{a} axis. VESTA 3 has been used for crystal structure visualization~\cite{Momma}.}
	\label{fig:fig1}
\end{figure}

\begin{figure*}[!t]
	\includegraphics[width=0.8\textwidth]{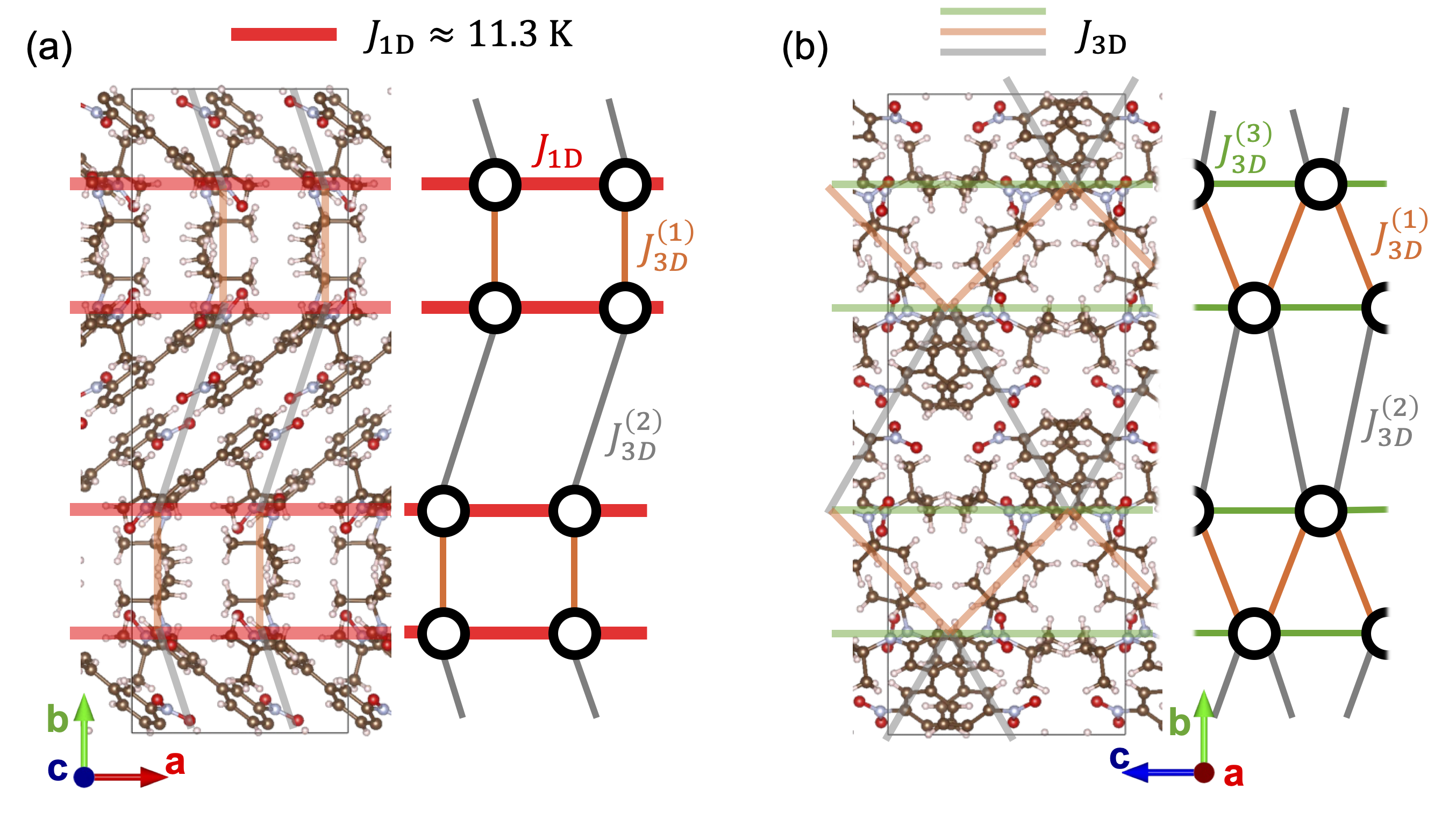}
	\caption{Single-crystal diffraction reveals a favorable stacking of BoNO molecules which creates a quasi-1D $S=1$ system. \blue{(a) The crystal structure in the $ab$ plane shows chain propagation along the crystallographic $a$ axis, overlaid with the dominant spin exchange path (red), corresponding to $J_{\text{1D}} / k_B \approx 11.3$~K. Weaker interchain couplings, $J_{\text{3D}}^{(1)}$ and $J_{\text{3D}}^{(2)}$, are indicated by orange and gray lines, respectively. (b) The structure in the $bc$ plane (the chains extend into the page) emphasizes the weaker interchain couplings $J_{\text{3D}}^{(i)}$ ($i = 1, 2, 3$) shown by orange, gray and green lines, respectively. Schematic representations of spin-exchange couplings are provided for clarity alongside each crystallographic projection.}}
	\label{fig:fig2}
\end{figure*}

The BoNO crystallographic data is summarized in Table \ref{tab:crystalographic_data}. The core molecular structure consists of two meta-positioned aminoxyl groups (Fig.~\ref{fig:fig1}), forming a ferromagnetic coupling unit~\cite{Rajca1994}. \blue{The} introduction of additional functional groups improves molecular stability and facilitates the formation of macroscopic single crystals.\\
\indent The introduced polar $-$NO$_2$ substituent aids the formation of a desired one-dimensional geometry due to the electrostatic interaction between the neighboring molecules. In the observed molecular chains, running along the crystallographic $a$ axis, electron-rich aromatic 3,5-bis($N$-$tert$-butylaminoxyl)benzene  moieties make an angle of $\approx \pm 45^\circ$ to the $a$~axis ~(Fig.~\ref{fig:fig1}~(b)). The stacked BoNO molecules are separated by $\textbf{a}/2$ along the chain, and $\pm \textbf{b}$ and $\pm \textbf{b}/2 \pm \textbf{c}/4$ between neighboring chains ($z = 6$)  in a distorted triangular (hexagonal) configuration, hence reducing frustration (Fig.~\ref{fig:fig2}). Each unit cell consists of 8 BoNO molecules, forming 4 parallel chains along the crystallographic $a$ axis. \textit{Ab initio} molecular orbital calculations for a single molecule, and DFT calculations for the unit cell, show that the $S=1$ spin density is primarily distributed between the two aminoxyl groups, with a significant overlap along the chain. The intramolecular  coupling was calculated to be $|J_\text{FM}|\blue{/k_B} \approx 450$~K, therefore substantially above room temperature.

Figure \ref{fig:fig2} shows the intrachain exchange coupling, $J_\text{1D}$, which results from the direct overlap between the $S=1$ orbitals of nearest neighboring BoNO molecules stacked along the $a$ axis. In contrast, the interchain $J_\text{3D}$ coupling is considerably smaller due to strong localization within C$-$H bonds in the large $tert$-butyl group which hinders the transferred exchange coupling between the neighboring chains. Considering only the nearest-neighbor interchain coupling, from the crystal structure we can distinguish three different coupling constants, $J_\text{3D}^{(i)}$, ($i=1, 2, 3$) which are located within the $bc$ plane. Unfortunately it was not possible to accurately calculate the size of $J_\text{3D}^{(i)}$ because the basic unit cell is quite large (416 atoms) and such calculations would require substantial computing power. Nevertheless, we can state that our low resolution calculation gave the average value $zJ_\text{3D} / J_\text{1D} \approx 0.02$.

\subsection{Static susceptibility}
The temperature-dependent static magnetic susceptibility ($\chi$) curves measured under an external applied magnetic field of 0.1~T both parallel ($a$ axis) and perpendicular ($b$ and $c$ axis) to the chain direction are shown in Fig.~\ref{fig:fig3}. For all orientations, data show almost identical behavior, namely a Curie-Weiss-like behavior at high temperatures with a pronounced, broad maximum at $T \approx 11$~K, characteristic for low-dimensional systems~\cite{Faridfar2022} with dominant antiferromagnetic interactions. The subsequent decrease in $\chi(T)$ is in line with an  $S=1$ antiferromagnetic chain model, where magnetic susceptibility vanishes at $T=0$~K with the gapped singlet ground state. The measured magnetic moment $\mu_\text{meas.} = (2.7 \pm 0.1)~\mu_B $ per formula unit (f.u.) is consistent with the $S=1$ system ($\mu_\text{theor.} = 2.82~\mu_B$/f.u.) and the Curie-Weiss temperature of $\Theta = -(23\pm4)$~K determine the antiferromagnetic nature of the intrachain interactions. The presented values for $\mu_\text{meas.}$ and $\Theta$ \blue{are} the average value measured for all three orientations. The low uncertainty depict the equivalency of the measured curves which shows that within the experimental error the observed magnetic susceptibility is isotropic. A precise determination of antiferromagnetic $J_\text{1D}$ coupling is done by fitting the closed-form expression derived by Padé-approximation of the numerical QMC calculations for an ideal Haldane $S=1$ model, calculated by Law et al.~\cite{Law2013}:

\begin{figure}[!t]
	\includegraphics[width=0.5\textwidth]{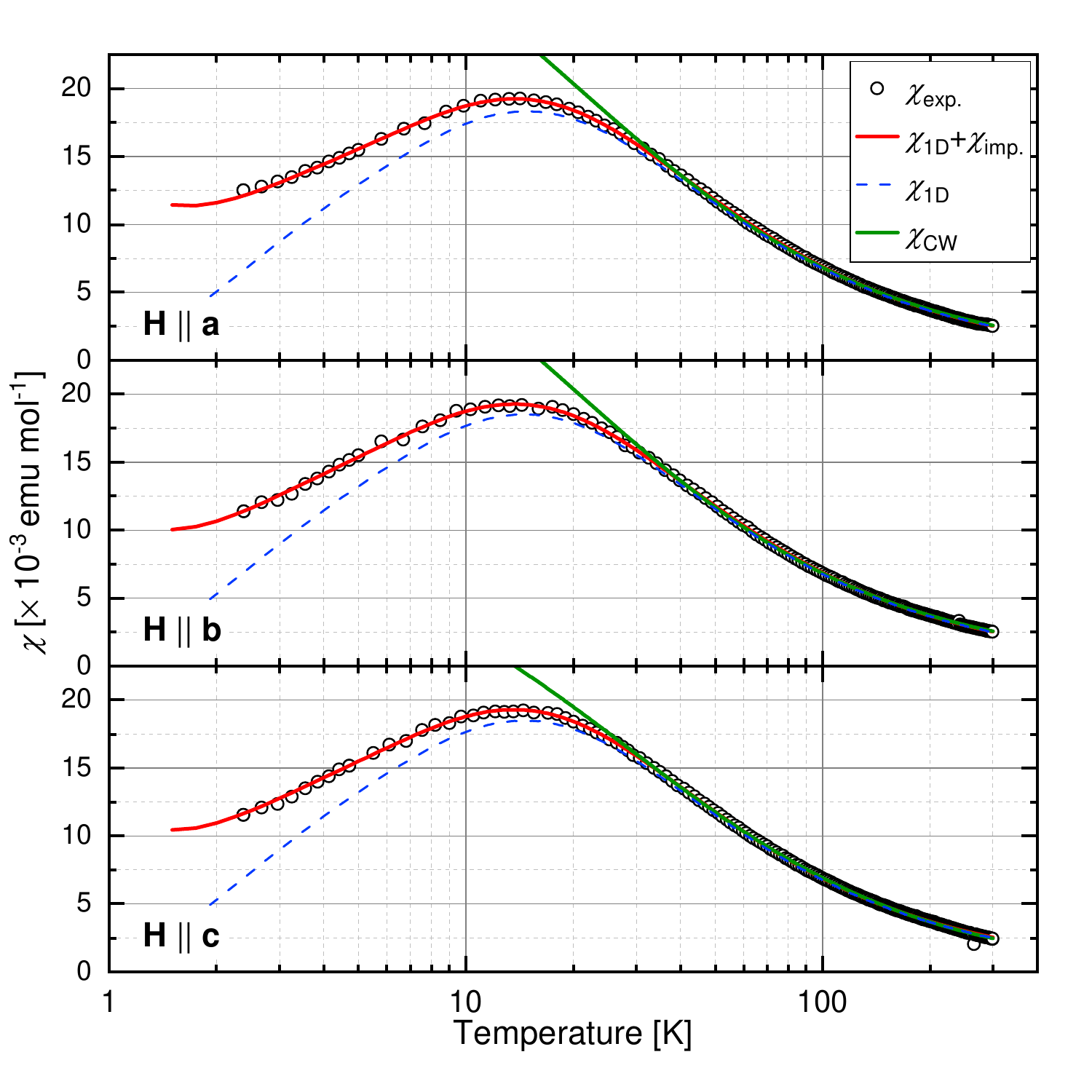}
	\caption{The low magnetic field measurements of spin susceptibility $\chi(T)$ in $\mu_0 H = 0.1$~T for $\textbf{H} \parallel \textbf{a}$ (along the chains), and $\textbf{H} \parallel \textbf{b},\textbf{c}$ (perpendicular to the chains) display a characteristic maximum consistent with a Haldane system with spin exchange constant $J_\text{1D}\blue{/k_B} = (11.3\pm0.1)$~K. The open circles are measured data, \blue{and }full red and dashed blue lines are fits to expression~(\ref{eq:Law}) \blue{(Ref.~\cite{Law2013})}, with and without the impurity contributions, respectively, in a wide temperature range. Note that in the impurity\blue{-}free  susceptibility (dashed blue) the maximum value is shifted to $\approx 15$~K. Curie-Weiss like behavior (green line) is observed down to $T \approx 30$~K.
	}
	\label{fig:fig3}
\end{figure}

\begin{align}
	\chi_\text{\blue{1D}}(T) =& x \chi^\text{(imp.)} + (1-x) \mathcal{F}_6 \left(\frac{J_\text{1D}}{\blue{k_B}T}\right)\chi_\text{C} e^{-\Delta/\blue{k_B}T}, \label{eq:Law}\\
\mathcal{F}_{n}\left(\frac{J_\text{1D}}{\blue{k_B}T}\right) &= \frac{1+\sum_{i=1}^{n}A_i\left(\frac{J_\text{1D}}{\blue{k_B}T}\right)^i}
{1+\sum_{i=1}^{n}B_j\left(\frac{J_\text{1D}}{\blue{k_B}T}\right)^i},
\end{align}
where $\chi^\text{imp.}=\frac{(g\mu_B)^2S(S+1)}{3k_B T}$ is the standard 3D Curie susceptibility for unpaired $S=1$ spins used to model paramagnetic impurity contribution, $x$ is the impurity concentration, $J_\text{1D}$ is the intrachain interaction, \blue{Haldane gap is given by} $\Delta = 0.411J_\text{1D}$, and $A_i$ and $B_j$ represent Padé coefficients in the susceptibility \blue{expression} for the $S=1$ system \footnote{\protect{\blue{Padé-approximation coefficients}: $A_1 = 0.679$, $A_2 = 1.270$, $A_3 = 0.655$, $A_4 = 0.141$, $A_5 = 0.088$, $B_1 = 1.600$, $B_2 = 2.653$, $B_3 = 2.516$, $B_4 = 1.678$, $B_5 = 0.420$, $B_6 = 0.041$}}. The paramagnetic impurity contribution is minute,  $x = (1.4 \pm 0.2) \%$, and can easily account for the slower reduction of susceptibility at low temperatures. The dominant chain term in Eq.~\ref{eq:Law} reduces to the paramagnetic Curie susceptibility in the high temperature limit, and the exponential term models the singlet-triplet gap. By removing the impurity contribution the maximum in susceptibility slightly shifts to $\approx 15$~K.  The fits to the experimental data yield an average value of $J_\text{1D}\blue{/k_B} = (11.3\pm0.1)$~K for all three orientations. This corresponds to the \blue{theoretical} gap value of $\Delta\blue{/k_B} = (4.6\pm0.4)$~K.  \blue{In reality, the}  gap \blue{is} reduced by the presence of weak $3$D interactions~\cite{Wierschem20142}, as well as in the case of monoradical and paramagnetic impurities present in the sample~\cite{Mukai1975}. Following the study of the thermodynamic properties of $S=1$ chain~\cite{Faridfar2022}, from the $J_\text{1D}$ value one can expect that the vanishing susceptibility is expected only at temperatures below $1$~K. \\ 
\indent The high-temperature values of susceptibility were analyzed (Appendix~A) to determine the intramolecular coupling value $|J_\text{FM}|$, and we find the data consistent with  $|J_\text{FM}|\blue{/k_B} \gtrsim 500$~K, in good agreement to the DFT result.\\  
\indent We can also include the effect of interchain interactions using a standard~\cite{Kahn} mean-field correction $\chi_\text{3D} = \chi_\text{1D} / [1-(zJ_\text{3D}/N_\text{A} g^2 \mu_B ^2)\chi_\text{1D}]$, where $\chi_\text{1D}$, $N_\text{A}$ and $\mu_B$ are the theoretical $S=1$ chain susceptibility modeled using  \blue{ the expression~(\ref{eq:Law})}, Avogadro's constant and Bohr magneton, respectively. The relative scale of $zJ_\text{3D}/J_\text{1D}$ makes it difficult to fit the absolute value \blue{of} $zJ_\text{3D}$ in \blue{the} paramagnetic phase. Nevertheless, the data conservatively suggest that the total interchain coupling $zJ_\text{3D}\blue{/k_B} \approx 0.6$~K is almost two orders of magnitude smaller than $J_\text{1D}$. 
\begin{figure}[!t]
	\includegraphics[width=0.5\textwidth]{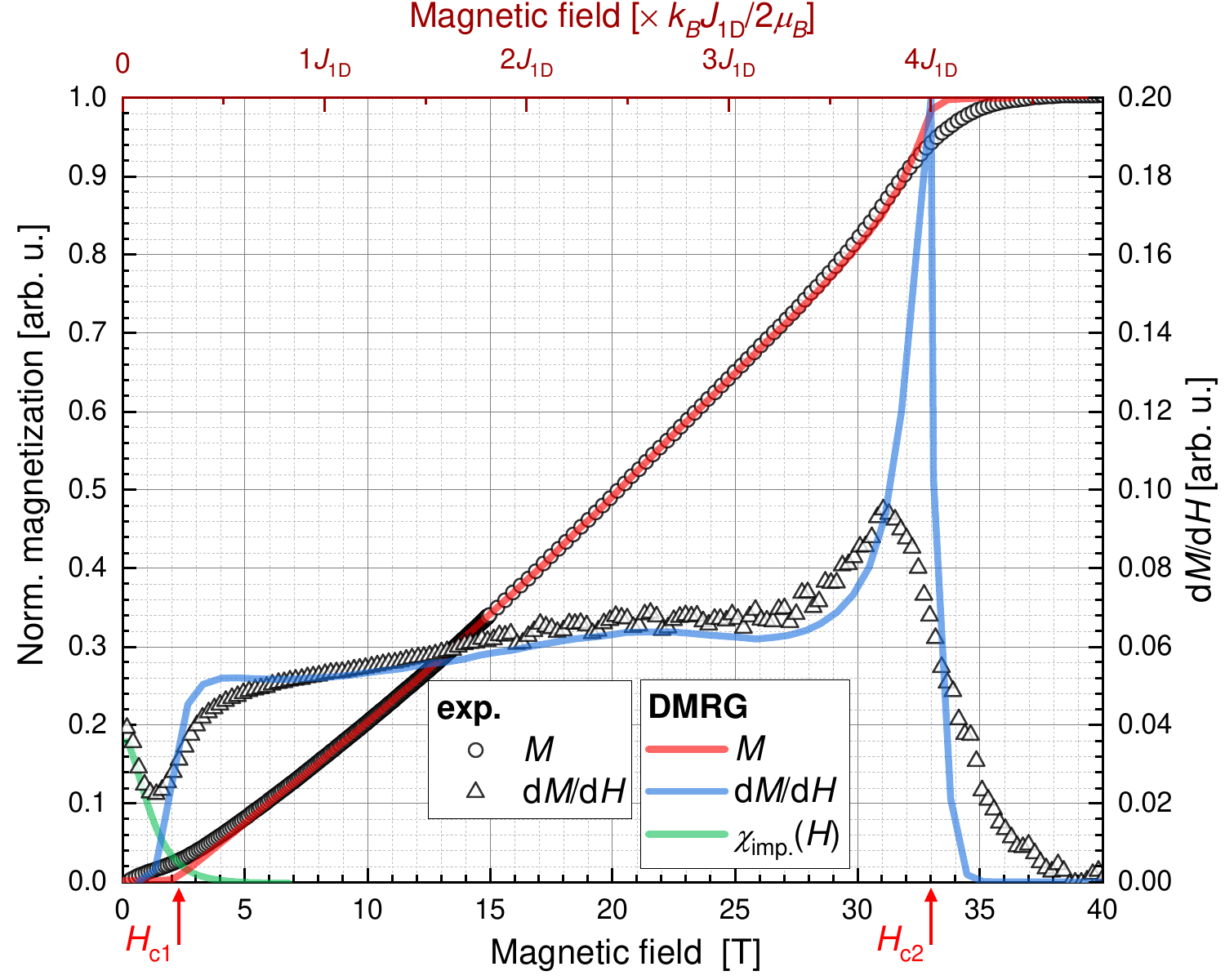}
	\caption{The high-field magnetization (black circles) measured at $T=1.3$~K, in pulsed magnetic fields up to $\mu_0 H = 40$~T agree well with the DMRG calculation of a prototypical Haldane system (red line) of tetragonal symmetry from \cite{Wierschem2014}, for $z J_{\text{3D}}/J_{\text{1D}} = 0.04$. \blue{Additionally, e}xperimental (triangles) and theoretical (blue line) data for $dM/dH$ also show a good agreement. \blue{The visible upturn in $dM/dH$ data at low magnetic fields arises from the paramagnetic impurities ($<1\%$) in the sample, modeled using the $S=1$ Brillouin function.}}
	\label{fig:fig4}
\end{figure} 

\subsection{High-field magnetization}
The high-field magnetization measurement at a temperature of $T=1.3$~K is shown in Fig.~\ref{fig:fig4}. We compare our magnetization and susceptibility data to a density-matrix renormalization group (DMRG) calculation from Ref.~\cite{Wierschem2014} that was done for a $S=1$ chain with $zJ_{\text{3D}}/J_{\text{1D}} = 0.04$ (with $z = 4$). As can be seen, the \blue{DMRG result matches the experiment}. It should be noted however, that \blue{the} exact value of the $zJ_{\text{3D}}/J_{\text{1D}}$ can be retrieved only once the values of $H_\text{c1}$ and $H_\text{c2}$ at $T = 0~\text{K}$ are determined.  The $H_\text{c1}$ ($H_\text{c2}$) value at 0~K is lower (higher) than that seen in Fig.~\ref{fig:fig4}, but nevertheless it gives us a good estimate of total interchain coupling. The magnetization curve displays a characteristic behavior at the lower critical field $\mu_0 H_\text{c1} \approx 2$~T, while close to the upper critical field $\mu_0 H_\text{c2} \approx  33$~T the field dependence is smeared due to the finite temperature effects. At intermediate magnetic fields, the increase in magnetization $M(H)$ is quasi-linear before reaching  complete polarization~\cite{Chitra1997}.  Although  comparison to the theoretical curve is not a fit, the used $z J_{\text{3D}}/ J_{\text{1D}} =0.04$ value is consistent \blue{with} our mean field result of 0.053 obtained from dc susceptibility.\\
\indent The measured  $H_\text{c1}$ ($H_\text{c2}$) values indicate that in this Haldane system the entire field-induced phase diagram can be thoroughly studied, which is not possible in known Haldane systems with low magnetic anisotropy.
\begin{figure}[t!]
	\includegraphics[width=0.5\textwidth]{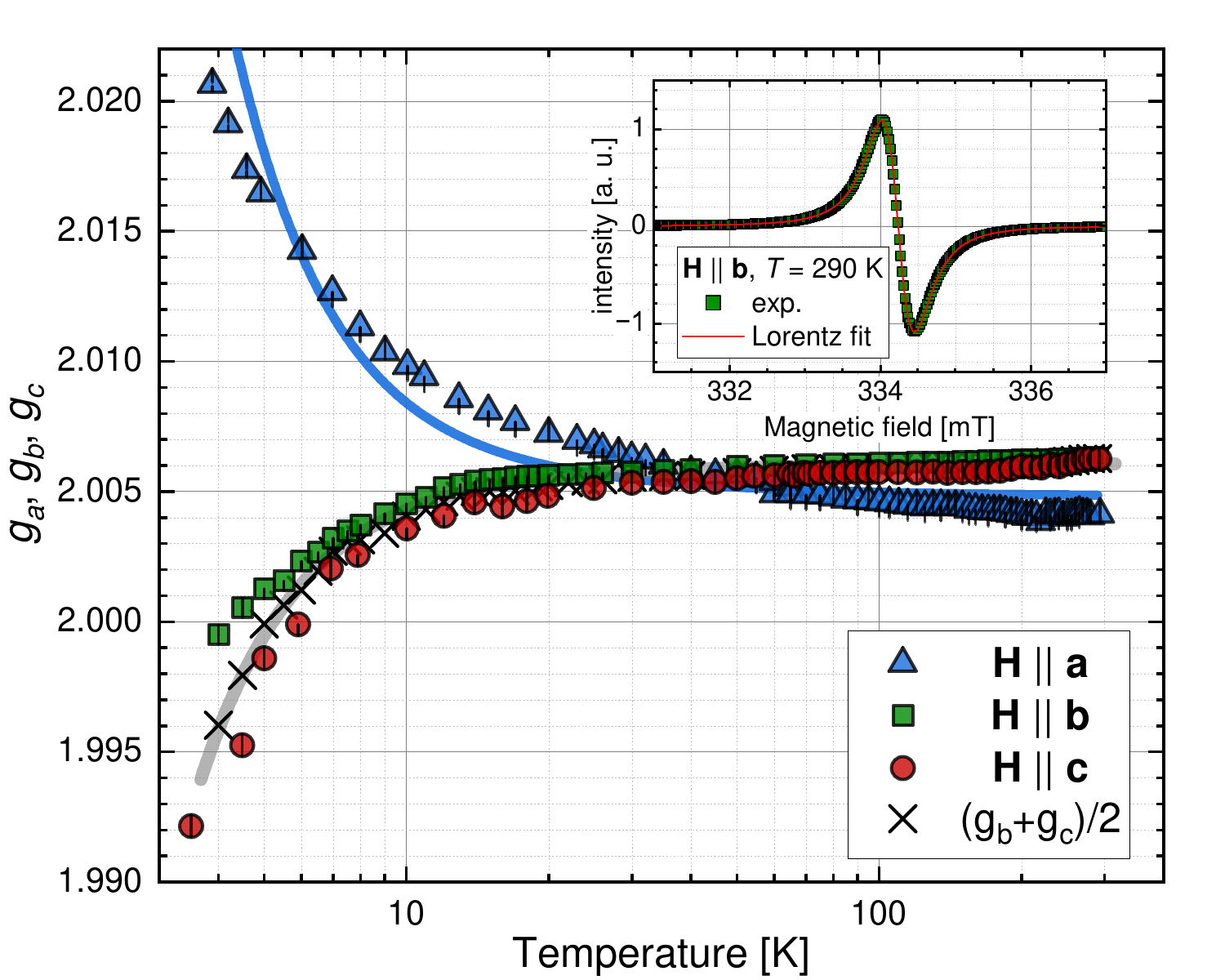}
	\caption{The values of $g_a$, $g_b$ and $g_c$ extracted from the lineshape fits of the X-band EPR spectra are plotted in a wide temperature range from $300$~K to $3.5$~K. The observed temperature dependence is successfully accounted for using Nagata et al. model~\cite{Nagata1972, Nagata1977} (blue and gray lines). Crosses mark the average values of the $g$ factor measured perpendicular to the chains $(g_b + g_c)/2$ for easier comparison to the model.  Below  10~K the system is no longer a paramagnet and deviations from the model are expected. \blue{The} inset shows \blue{the} EPR line measured for $\textbf{H} \parallel \textbf{b}$ at $290$~K and the corresponding Lorentz fit.}
	\label{fig:fig5}
\end{figure}
\subsection{Electron paramagnetic resonance}
\indent An X-band electron paramagnetic resonance (EPR) measurements  used to confirm the (nearly) isotropic spin Hamiltonian of the BoNO molecule (Fig.~\ref{fig:fig5}). The single crystal EPR spectra were measured at room temperature for the magnetic field oriented parallel and perpendicular to the chain direction. A single narrow EPR spectral line (Fig.~\ref{fig:fig5} inset) is observed for each orientation and Lorentzian fits to the data give $g_a = 2.0041$ and $\mu_0\Delta H_a = 2.99$~mT for the field along the chains, and $g_b = 2.0065$,  $\mu_0\Delta H_b = 0.75$~mT, $g_c = 2.0063$ and $\mu_0\Delta H_c = 0.89$~mT for the fields perpendicular to the chain direction. Remarkably small $g$ factor deviations from \blue{the} free electron value $g_e = 2.00232$ and narrow linewidths are consistent with the light-element nature of BoNO molecules where spin-orbit coupling and the resulting magnetic anisotropic Hamiltonian terms are very small.\\
\begin{figure}[!t]
	\includegraphics[width=0.5\textwidth]{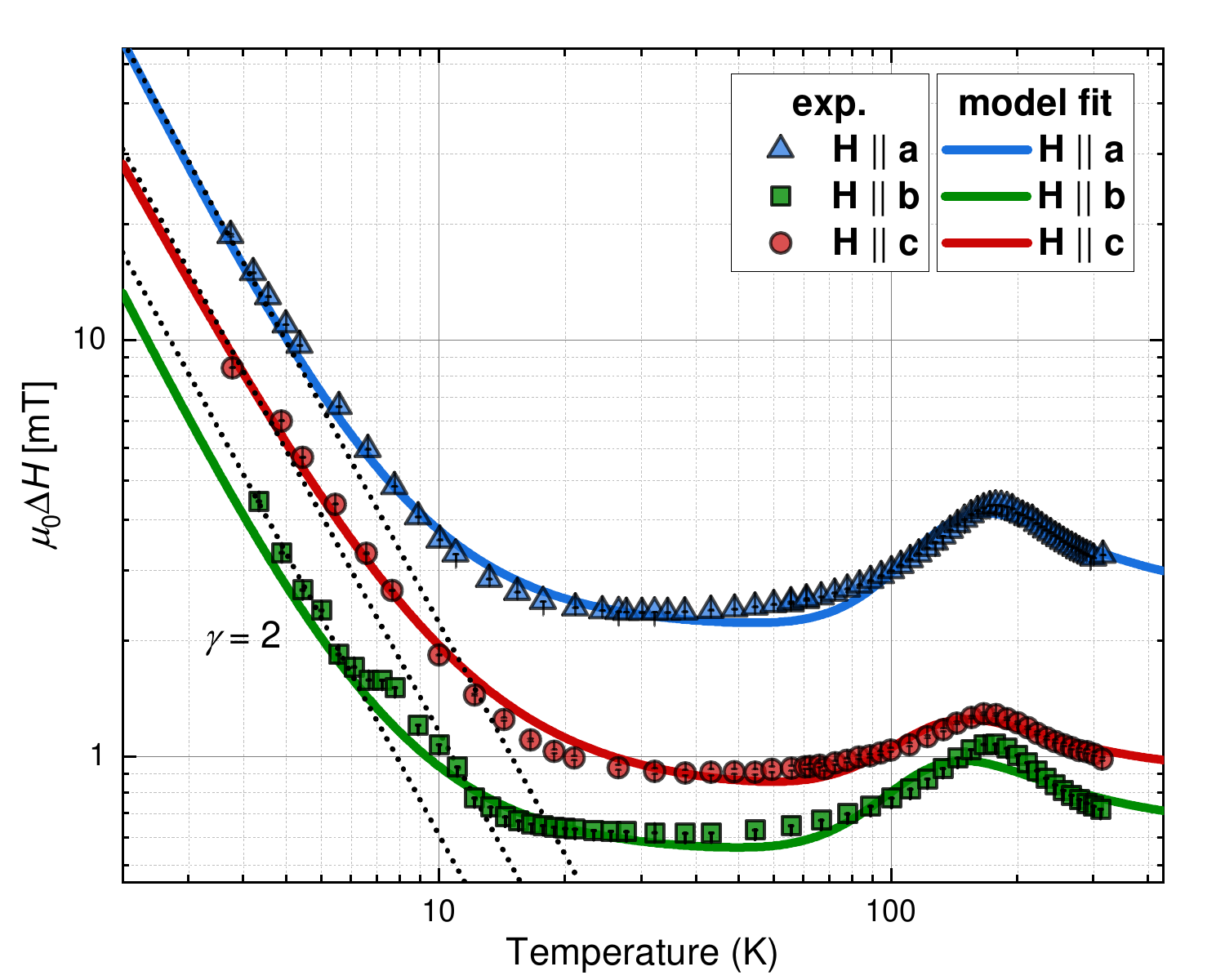}
	\caption{Temperature dependence of the X-band EPR linewidth $\Delta H$ for magnetic field along the chain direction ($a$ axis, triangles) and perpendicular to it ($b$ and $c$ axis, squares and circles, respectively). The data \blue{shows a good agreement with} the proposed model (\ref{eq:deltaHT}) with both BPP and power-law terms. Dotted black lines illustrate $\propto T^{-2}$ behavior observed in systems with short-range correlations.}
	\label{fig:fig6}
\end{figure}
\indent Temperature dependence of the $g$ factor is most pronounced at low temperatures where the $g$ factor change reaches about $|\Delta g| = 0.015$. The change is positive for the magnetic field aligned along the chain $a$ axis, while it is negative for the other two orientations (Fig.~\ref{fig:fig5}). This known effect is expected to arise from emerging short-range fluctuations along the chains, which shift the EPR \blue{spectral line}~\cite{Oshikawa1999, Oshikawa2002, Ohta2023, Herak2011}. We used a nearest-neighbor (NN) uniaxial Heisenberg chain model in  a paramagnetic state developed by Nagata et al.~\cite{Nagata1972, Nagata1977} to quantitatively address the temperature dependencies of the effective $g$ factors:
\begin{align}
	g_{\parallel} ^\text{eff.} (T)&= g_\parallel + \frac{6 \alpha g_\parallel}{10x(T)} \left[ \frac{2+u(T)x(T)}{2-u(T)^2}-\frac{2}{3x(T)}\right], \\
	g_{\perp} ^\text{eff.} (T)&= g_\perp - \frac{3 \alpha g_\perp}{10x(T)} \left[ \frac{2+u(T)x(T)}{2-u(T)^2}-\frac{2}{3x(T)}\right],
\end{align}
where $g_{\parallel} ^\text{eff.}$ and $g_{\perp} ^\text{eff.}$ stand for the  temperature-dependent effective $g$ factors measured for \blue{the} magnetic field along and perpendicular to the chains, respectively. The parameter $u(T) = \coth[1/x(T)]-x(T)$, where $1/x(T) = J_\text{1D} S(S+1)/k_B T$ is related to the spin correlation function, and $\alpha = - \mu_0 g^2 \mu_B^2/(4 \pi J_\text{1D}r_\text{NN}^3$) models the dipolar magnetic interaction between $S=1$ sites separated by $r_\text{NN}$ along the chain.  Since our spin chain environment is not uniaxial, we compare the $	g_{\perp} ^\text{eff.} (T)$ dependence to the average $g$ factor  perpendicular to the chains, $(g_b + g_c)/2$. By using the previously determined $J_\text{1D}$, $g_{\parallel}$ and $g_{\perp}$, and  $\alpha$ as the only free parameter the model fits our data (Fig.~\ref{fig:fig5}) for \blue{$\alpha = 1.7\cdot10^{-4}$, i.e.} $r_\text{NN} = 11$~\AA. Some deviations are visible in $g_a (T)$ for $\blue{k_B}T \lesssim J_\text{1D}$, but this is expected since there the quantum effects become dominant while the model is classical in nature. The fitted $r_\text{NN}$ value is larger than the mean intrachain distance between molecules ($\approx5~$\AA), however this can be explained by the  \blue{simplicity of the model} - in reality dipolar contribution has \blue{a} longer range than only NN, and in BoNO the spin wave function is distributed via several atoms of the molecule, which is unaccounted for by the model.\\
\indent The presence of short-range fluctuations for $\blue{k_B}T \leq J_\text{1D} $ is further corroborated by \blue{the} temperature dependence of the fitted linewidths $\Delta H$ (Fig.~\ref{fig:fig6}), which we model using the expression:
\begin{equation}
	\Delta H(T) = \Delta H_0 + a \frac{\tau_c}{1+(\omega_\text{EPR} \tau_c)^2} + b T^{-\gamma}.
	\label{eq:deltaHT}
\end{equation}
\begin{figure}[!t]
	\includegraphics[width=0.5\textwidth]{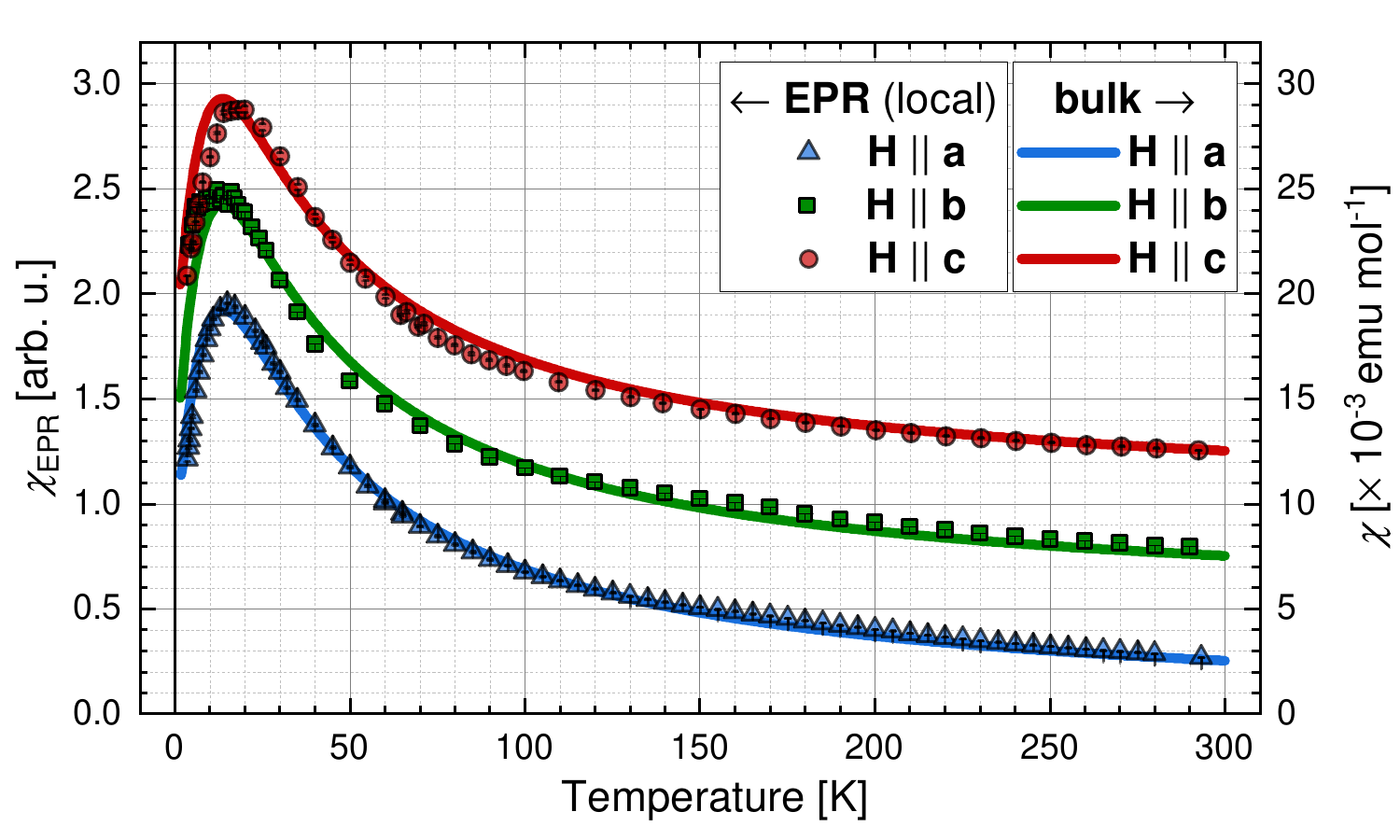}
	\caption{The EPR susceptibility data at $\mu_0 H=334$~mT, \blue{obtained by a double-integration of the derivative of the EPR spectrum (inset of Fig.~\ref{fig:fig5})}, is \blue{compared} to the corresponding bulk measurements (each dataset is offset by 0.5 [\blue{arb}.u.], i.e. 0.005 emu/mol, for clarity).}
	\label{fig:fig7}
\end{figure}
 In addition to the temperature-independent intrinsic linewidth $\Delta H_0$, the observed high-temperature maximum at $T=150$~K is treated using Bloembergen-Purcell-Pound (BPP) type of relaxation mechanism with a well-defined frequency of the local magnetic field fluctuations, and the last term describes short-range correlations~\cite{Oshikawa2002,Furuya2015} characterized by a power-law exponent $\gamma$. The BPP-modeled fluctuations are probably the result of the molecular librations and rotations defined by the energy scale $E_a$ and characteristic correlation time $\tau_c$, transferred via intramolecular dipolar coupling between $S=1$ spins and the nearby hydrogen magnetic moments. We can exclude any structural transition since the crystal structure measured at 300~K and 100~K revealed only a contraction of the unit cell. The BPP correlation time $\tau_c$ is usually a thermally activated term $\tau_c=\tau_\infty \exp(E_a/k_B T)$. The extremal relaxation rate is expected for $\omega_\text{EPR} \tau_c=1$, where the observed linewidth is maximal, with $\omega_\text{EPR}$ as the measurement frequency. The pronounced low-temperature broadening we identify as short-range order effects give  a characteristic power-law behavior with exponent $\gamma$.
The fitted curves yield $\tau_\infty  = 7.5 \cdot10^{-13}$~s, $E_a /k_B = 575$~K and $\gamma = 2.2$. The magnitude of the fitted activation energy ($E_a = 4.2$~kJ/mol) matches the expected range for the energy barriers between different molecular conformations~\cite{Durig1974}. The low-temperature increase of $\Delta H(T)$ approximately follows $~T^{-2}$ power-law dependence, which is usually associated with the short-range correlations in quantum spin systems with the staggered field.\\
\indent We complement our dc susceptibility data with the measurements of EPR susceptibility (Fig.~\ref{fig:fig7}) at the external magnetic field of $\mu_0 H=334$~mT. The data follow a Curie-Weiss dependence down to a maximum  value at $\approx15$~K, and it is suppressed on further cooling due to the singlet-triplet gap. In the figure we compare the measured EPR susceptibility to the impurity-free data of Fig.~\ref{fig:fig3}. As can be seen, the datasets match \blue{to} each other, confirming the validity of the Curie-term subtraction.\\ 
\begin{figure}[t!]
	\includegraphics[width=0.5\textwidth]{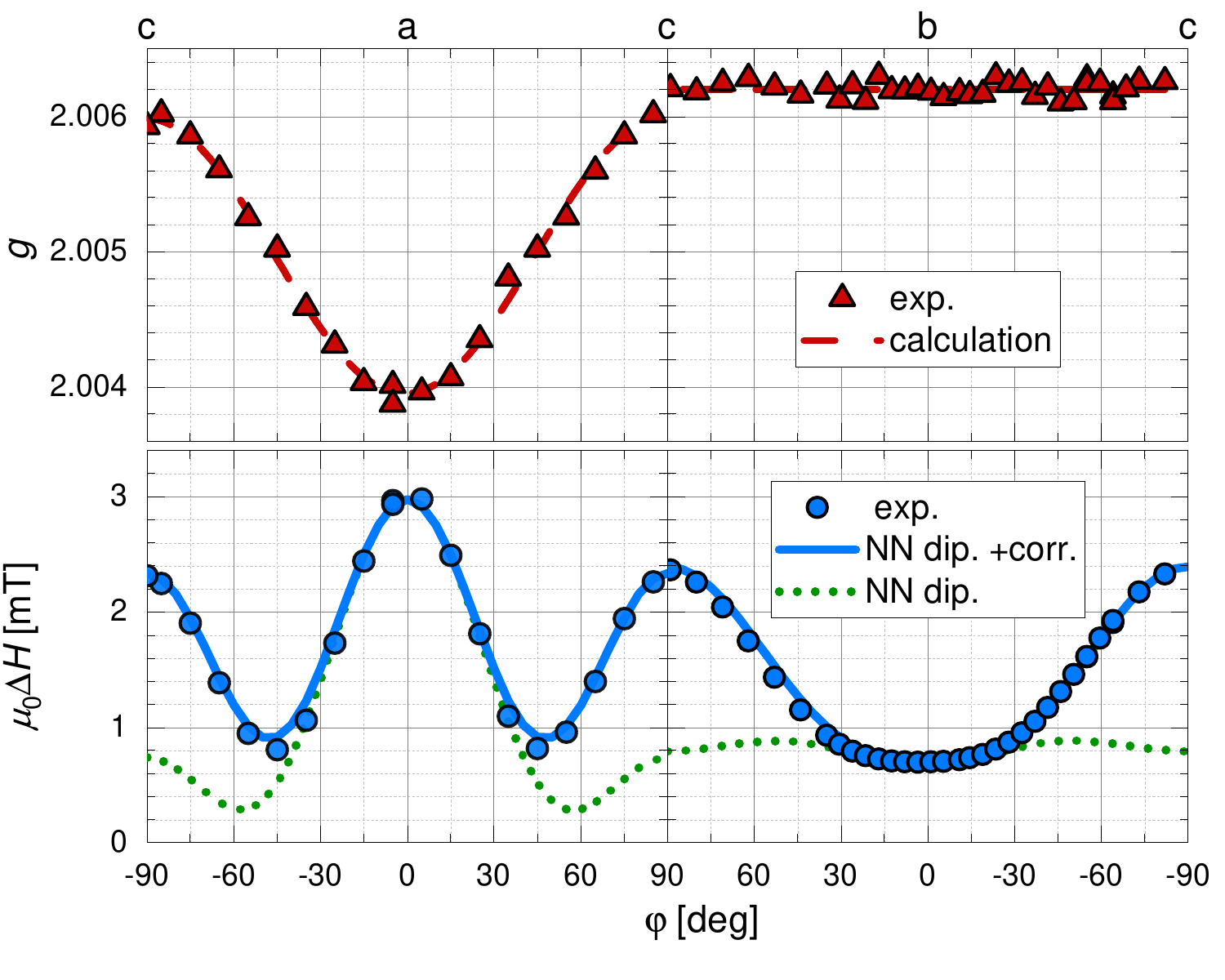}
	\caption{The angular dependence of \blue{the} $g$  factor and linewidth $\Delta H$ values taken along $c-a-c-b-c$ path. The change in $g$ factor is attributed to a small molecular orbital anisotropy while $\Delta H$ values are explained by the nearest-neighbor dipolar interaction between $S=1$ coupled units. The red dashed line is calculated by rotating the $g$ tensor, using the previously determined $g_a$, $g_b$ and $g_c$ values.  The green dotted lines represent the calculated NN dipolar broadening, and the magnitude is corrected when small intramolecular interactions between $S=1/2$ spins are taken into account (full lines).}
	\label{fig:fig8}
\end{figure} 
\indent Finally, we discuss the angular dependence of the EPR spectra taken at $T=300$~K shown in  Fig.~\ref{fig:fig8}. In the upper panels we plot the $g$ factors against the angle measured away from the chain direction in the crystal $ac$ plane (left panel) and in the $bc$ plane that is perpendicular to the chains (right panel). The data show that the $g$ factor is almost angle independent in the plane perpendicular to the chain direction. The main angular dependence is thus with respect to the angle from the chain direction. The fits to the angular dependencies are generated by a simple projection of the diagonal $g$ tensor with principle values corresponding to the previously determined $g_a$, $g_b$ and $g_c$ values.\\
\indent The angular dependence of the EPR linewidth, shown in the lower panel of Fig.~\ref{fig:fig8}, holds the information about the magnetic anisotropy. For the field rotation in the $ac$ plane, the linewidth $\Delta H$ shows a minimum value at $\varphi \approx 50^\circ$ and maximum for the field along the chain direction - a characteristic behavior of quasi-1D antiferromagnets~\cite{Bencini}. As the spin-orbit coupling is very weak, we can assume that the main magnetic anisotropy is the dipolar interaction between neighboring $S=1$ molecular sites. The change in EPR linewidth due to the magnetic dipolar interactions can be calculated in the exchange narrowing limit:
\begin{equation}
	\Delta H = C\frac{M^\text{dip.}_2}{J_\text{1D}},
	\label{eq:deltaHepr}
\end{equation}
where $M_2^\text{dip}$ is the second moment of the resonance due to the dipolar interactions, and lineshape-dependent constant $C$ ranges between 1-2 in similar systems~\cite{Gulley1970, VanVleck1948}. In the point-spin model where the effective spin $S=1$ is centered between the radical NO groups, the $M_2^\text{dip.}$ is easily calculated: 
\begin{equation}
	M_2^\text{dip.} = \frac{3}{4}S(S+1) \frac{\mu_0^2 \mu_B^4 g^4}{(4 \pi)^2} \sum_{i \in \text{NN}}\frac{(3\cos^2 \Theta_i -1)^2}{r_i^6},
	\label{eq:M2}
\end{equation}
where $\Theta_i$ is the angle between the external magnetic field and the vector $\textbf{r}_i = r_i \hat{\textbf{r}}_i$ that connects $S=1$ sites of the neighboring molecules. The sum converges rapidly due to \blue{the} $1/r_i^6$ term, so only the nearest neighbors have to be accounted for. The calculated $\Delta H$ with $C = 1.6$ and $J_\text{1D}\blue{/k_B} = 11.3$~K  qualitatively explains the observed data, as is shown by the dotted line in the lower panel of Fig.~\ref{fig:fig8}, even with such a simple approach. Reproducing the exact $\Delta H$ angle dependence quantitatively is challenging since there may be other contributing factors too~\cite{McGregor}. In our case, $\Delta H$ is further affected by the on-site (anisotropic) distribution of the molecular orbital, which requires substantial computer power to take into account.  Since spin-orbit coupling  is for such  \blue{a} light-element molecular system expected to be small, the single-ion anisotropy should be negligible in BoNO. Yet, \blue{an} extra term can still arise due to the intramolecular interaction of two $S=1/2$ spins forming the single $S=1$ unit. Therefore, in the last step of our analysis we add \blue{$A\sin^4 \varphi$} term in addition to \blue{the} dipolar contributions from Eqs.~(\ref{eq:deltaHepr}) and (\ref{eq:M2}). \blue{A good agreement to our data} is finally obtained with $A = 1.6$~mT.

\subsection{Nuclear magnetic resonance}
The nuclear magnetic resonance (NMR) spectral and relaxation measurements have  previously frequently been used in studies of similar compounds to confirm and progressively improve the theoretical description of phenomena in low-dimensional quantum magnets~\cite{Klanjsek2008,Orignac2007,Blinder2017,Dupont2016,Dupont2018,Horvatic2020}. The \blue{hydrogen ($^1$H)} spectrum can be modeled by a simple spin $I=1/2$ NMR Hamiltonian:
\begin{align}
	\mathcal{H} &= {^1}\gamma  (1 + \mathbf{K}) \textbf{B} \cdot \textbf{I},
	\label{eq:NMRHam}
\end{align}
where  ${^1}\gamma/2\pi = 42.5774$~MHz/T is the corresponding gyromagnetic constant, $\textbf{I}$ is the nuclear spin operator, $\textbf{B}=\mu_0 \textbf{H}$ is the external magnetic field and $\mathbf{K} $ is the nuclear shift tensor. The spectral line observed at a given frequency $f_\text{NMR} = {^1}\gamma |B| \left\{1 + K(\theta, \phi)\right\}$ has a frequency shift $K(\theta, \phi)$ correlated to the respective site properties and the external magnetic field orientation defined by the spherical angles ($\theta$, $\phi$), where $\theta = 0^\circ$, $\phi = 0^\circ$ corresponds to the magnetic field aligned with the sample's crystallographic $c$ axis. The $I=1/2$ NMR Hamiltonian is coordinate inversion-symmetric, and consequently, for an arbitrary magnetic field direction, four spectral lines are expected for \textit{each} crystallographic $^1$H site within the \textit{Pbca} unit cell. However, fewer spectral lines will be observed for the magnetic field directions along the crystal symmetry axes. 

\begin{figure}[t!]
	\includegraphics[width=0.5\textwidth]{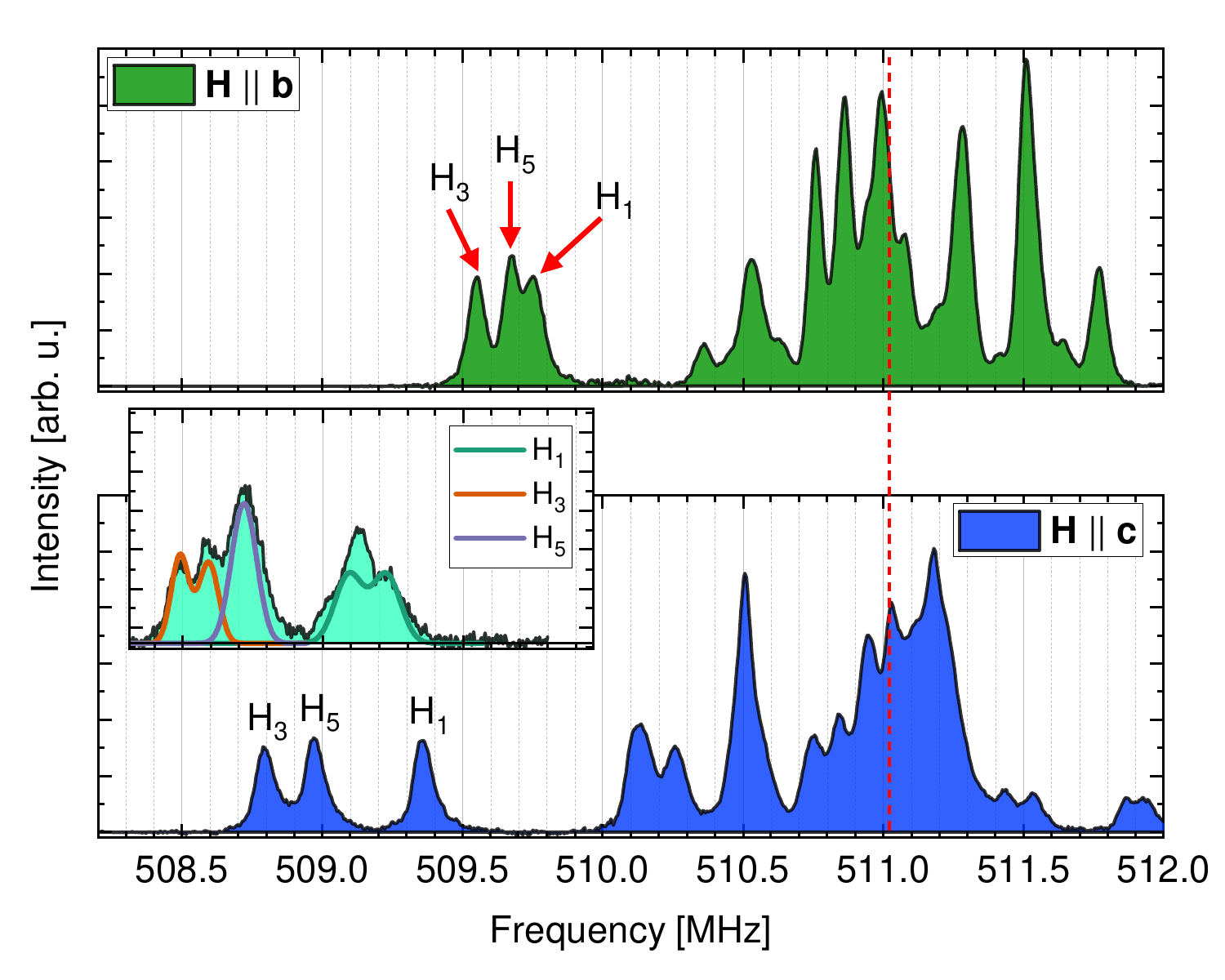}
	\caption{The frequency-swept $^1$H NMR spectra of BoNO single crystal at  $\mu_0 H=12.002$~T, along the crystalline $b$ and $c$ axes are shown in the upper and lower panel, respectively. The spectra are measured at 25~K. Most of the spectrum is concentrated close to $f_0 = 511.033$~MHz  (red dashed line) and cannot be resolved and assigned to the respective $^1$H atoms. However, we have successfully assigned the spectral lines with the strongest hyperfine nuclear shifts to the H$_1$, H$_3$ and H$_5$ crystallographic positions, respectively. These sites create three lines for $\textbf{H}\parallel \textbf{b}$ and $\textbf{H} \parallel \textbf{c}$, but split for $\textbf{H}$ away from crystallographic axes, as seen in the inset. Besides the measured NMR spectrum, \blue{the} inset also shows the calculated positions corresponding to these sites (see the main text).}
	\label{fig:BoNOspectra12T}
\end{figure}

\blue{The} $^1$H NMR spectrum \blue{of BoNO} at $\mu_0 H=12.002$~T in the paramagnetic phase for magnetic field orientations along the high-symmetry crystal axes $\textbf{H}\parallel \textbf{b}$ and $\textbf{H} \parallel \textbf{c}$ are displayed in upper and lower panel of Fig.~\ref{fig:BoNOspectra12T}, respectively. Here, due to the strong electron-rich character of the aromatic rings, several $^1$H spectral lines are strongly shifted from the bare nucleus Larmor frequency $f_0 = {^1}\gamma B$.
The $^1$H NMR spectrum has two contrasting features: a set of high-intensity spectral lines spanning a relatively wide frequency range close to the bare nucleus frequency $f_0 = 511.033$~MHz that correspond to $^1$H sites with small shifts, and a region of well-defined spectral lines with considerable negative frequency shifts (marked in Fig.~\ref{fig:BoNOspectra12T} as H$_1$, H$_3$ and H$_5$). As mentioned earlier, for \blue{the} magnetic field along the $b$ and $c$ axis there are only three lines corresponding to the H$_1$, H$_3$ and H$_5$ sites, and when the sample is oriented away from the axes the lines split (inset of Fig.~\ref{fig:BoNOspectra12T}). The lines are assigned to the corresponding $^1$H sites shown in Fig.~\ref{fig:fig1}, through a procedure which we will elaborate \blue{on} later.  Although some of the high-intensity lines close to $f_0$ can be clearly distinguished, the frequent overlapping during the sample rotation in an external magnetic field or temperature dependence severely hinders the possibility of spectral line assignation. Furthermore, we argue that spectral lines close to $f_0$ are only marginally influenced by the $S=1$ system, so we will focus our analysis \blue{on} the set of lines with large hyperfine shifts.

\begin{figure}[!t]
	\includegraphics[width=0.5\textwidth ,trim={0mm 0mm 0mm 15mm},clip]{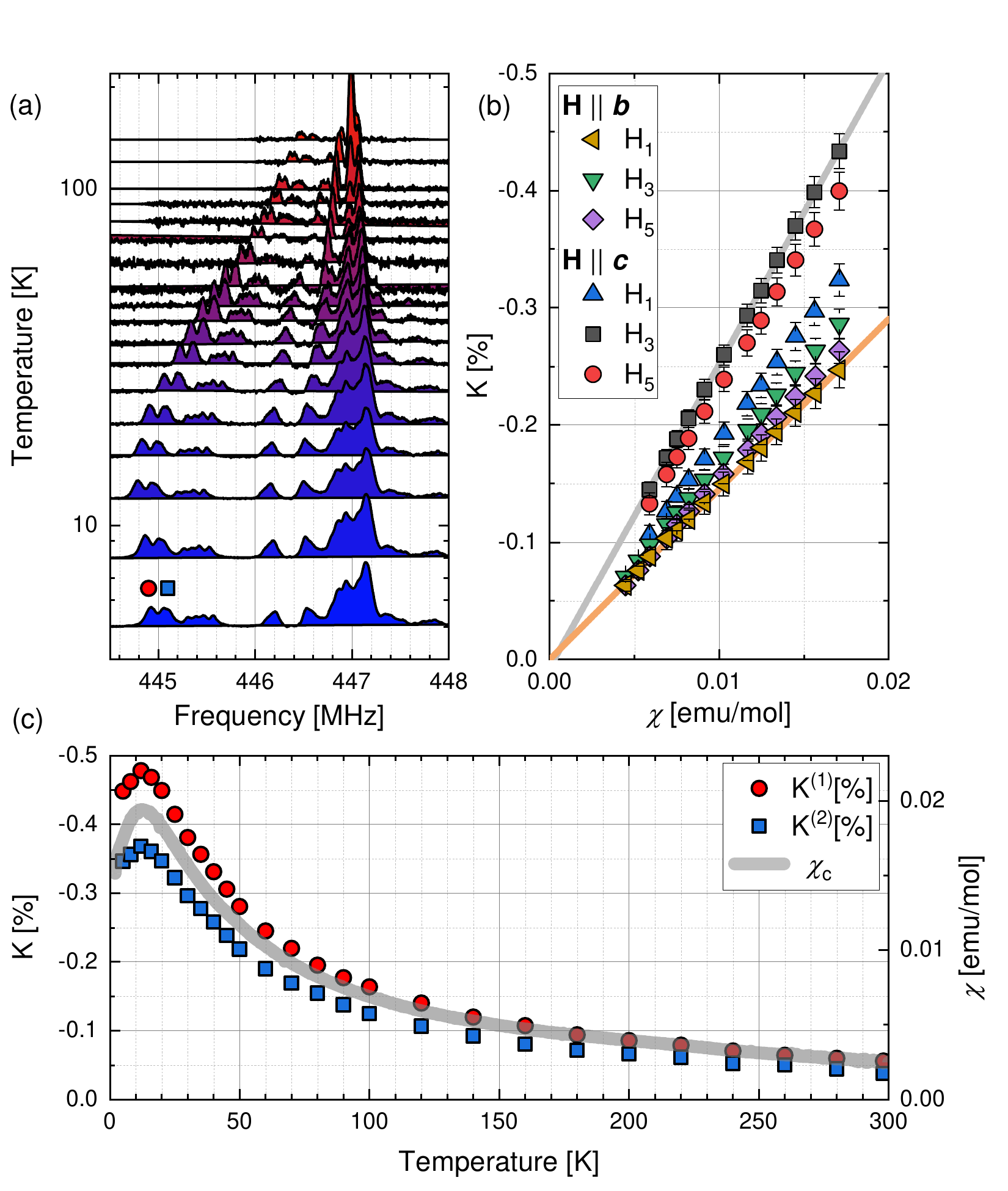}
	\caption{\blue{(a)} The temperature dependence of the NMR spectra for $\mu_0H = 10.5$~T, oriented at $\angle(H,c) = 45^\circ$ where the observed frequency shift is the largest. \blue{(b)} The $K-\chi$ plot for H$_1$, H$_3$ and H$_5$ protons, \blue{with} $\textbf{H}\parallel \textbf{b}$, $\textbf{H} \parallel \textbf{c}$, \blue{reveals} negligible orbital contribution.) \blue(c) The spectra are overlayed with \blue{the} scaled bulk susceptibility, \blue{represented by} a thick gray line.
	}
	\label{fig:fig10}
\end{figure}
\begin{table*}[ht!]
	\begin{tabularx}{16.5cm}{C|RRR|RRR|RRR}
		\multirow{2}{*}{component} & \multicolumn{3}{c|}{site \text{H}$_1$}& \multicolumn{3}{c|}{site \text{H}$_3$}& \multicolumn{3}{c}{site \text{H}$_5$} \\ 
		& \multicolumn{1}{c}{$\mathbf{K}$} & \multicolumn{1}{c}{$\mathbf{K}_\text{dip.}$} & \multicolumn{1}{c|}{$\mathbf{K}_\text{tr.}$} & \multicolumn{1}{c}{$\mathbf{K}$} & \multicolumn{1}{c}{$\mathbf{K}_\text{dip.}$} & \multicolumn{1}{c|}{$\mathbf{K}_\text{tr.}$} & \multicolumn{1}{c}{$\mathbf{K}$} & \multicolumn{1}{c}{$\mathbf{K}_\text{dip.}$} & \multicolumn{1}{c}{$\mathbf{K}_\text{tr.}$} \\ \hline \hline
		$xx$ & -3.952 & 0.049 & -3.998 & -3.011 & 0.229 & -3.237 & -8.314 & 0.105 & -8.416 \\
		$yy$ & -2.677 & 0.004 & -2.653 & -4.265 & 0.050 & -4.287 & -2.997 & -0.059 & -2.91 \\
		$zz$ & -3.866 & -0.053 & -3.762 & -5.177 & -0.279 & -4.848 & -4.658 & -0.045 & -4.561 \\ \hline
		$yz$ & -0.960 & -0.030 & -0.93 & 0.76 & 0.064 & 0.111 & -0.250 & -0.096 & -0.154 \\
		$xz$ & 0.005 & -0.005 & 0.010 & -0.702 & -0.028 & -0.674 & 0. & -0.075 & 0.075 \\
		$xy$ & -0.269 & 0.080 & -0.348 & -0.488 & 0.247 & -0.734 & 0. & 0.091 & -0.091
	\end{tabularx}
	\caption{Nuclear magnetic shift tensors (in \%) for the assigned $^1$H spectral lines. The shift tensor is dominated by the large diagonal elements, visually separated by a single line from the smaller, off\blue{-}diagonal elements, which are noticeably changed by the dipolar and demagnetizing contributions.}
	\label{table}
\end{table*}
The temperature dependence of the NMR spectra has been measured in a paramagnetic phase in an external field of $\mu_0 H = 10.5$~T with the field at an angle $\angle(H,c) = 45^\circ$ from $c$ to $b$ axis (shown in Fig.~\ref{fig:fig10}~(a)), where the H$_1$, H$_3$ and H$_5$ spectrum spans the broadest frequency range. At  this orientation, the spectrum becomes progressively wider as \blue{the} temperature is lowered, down to the temperature $T=16$~K\blue{,} where the maximal frequency shift is observed. The measurements were also done for $\textbf{H}\parallel \textbf{b}$ and $\textbf{H} \parallel \textbf{c}$, but we do not show these spectra here for clarity and brevity of the presentation. The paramagnetic NMR frequency shift reflects the bulk susceptibility \blue{of the sample} scaled to each spectral line and through the $K-\chi$ plot (Fig.~\ref{fig:fig10}~(b)) this relation can be used to determine the diagonal components of the shift tensor. The figure also shows that there is no anomalous behavior of $K$ in the entire paramagnetic temperature region, notably close to the BPP behavior seen by EPR (Fig.~\ref{fig:fig6}). Although no structural change has been seen by X-ray diffraction, since it is not sensitive to hydrogen positions, we wanted to corroborate it by characterizing \blue{$^1$H} NMR spectra. To show the $K(T)$ dependence in more detail we selected the $^1$H lines with the largest hyperfine coupling, i.e. the largest NMR shift (marked by symbols 	in Fig.~\ref{fig:fig10}~(a)), and plot the corresponding $K(T)$ dependence in Fig.~\ref{fig:fig10}~(c). The data follow the susceptibility curve, which confirms that the EPR line broadening with a maximum at $T = 150$~K is a dynamical effect accounted for by Eq.~(\ref{eq:deltaHT}) and not related to \blue{the crystal} structure.\\   
\indent A rigorous treatment of the NMR spectra and a precise determination the $\mathbf{K}$ shift tensor for multiple $^1$H sites is essential for the study of the microscopic properties \blue{of the system}. 
Here, we have performed a set of NMR spectral measurements at $T = 20$~K, $H=10.5$~T for different magnetic field orientations in $ac$ and $bc$ plane (Fig.~\ref{fig:rotation}~(a)-(c)). The \blue{positions of the spectral lines} were fit with the Gaussian lineshape, and subsequently fitted to expression for $f_\text{NMR} $. Quite generally, the $\mathbf{K}$ tensor can be expressed by the following equation~\cite{Nawa2013}:
\begin{equation}
	\mathbf{K} =- 4\pi\mathbf{N}\frac{\chi}{N_\text{A} v} +\mathbf{K}_\text{tr.} + \mathbf{K}_\text{dip.},
\end{equation}
where $\mathbf{N} = (0.038, 0.336, 0.626)$ stands for a diagonal demagnetization tensor calculated for a rectangular sample shape~\cite{Aharoni1998} with molar volume $v$. $\mathbf{K}_\text{tr.}$ and $\mathbf{K}_\text{dip.}$ are $\mathbf{K}$ tensor components due to transferred and dipolar hyperfine interactions. $\mathbf{K}_\text{tr.} = \mathbb{A}_\text{tr.} \chi/N_A \mu_B $, where $\mathbb{A}_\text{tr.}$ is the transferred hyperfine coupling tensor, while $\mathbf{K}_\text{dip.} = \mathbb{D} \chi/N_A \mu_B $, where $\mathbb{D}$ is the dipolar hyperfine coupling tensor.  $\mathbb{D}$  is easily calculated as:
\begin{figure}[!b]
	\includegraphics[width=1.01\columnwidth]{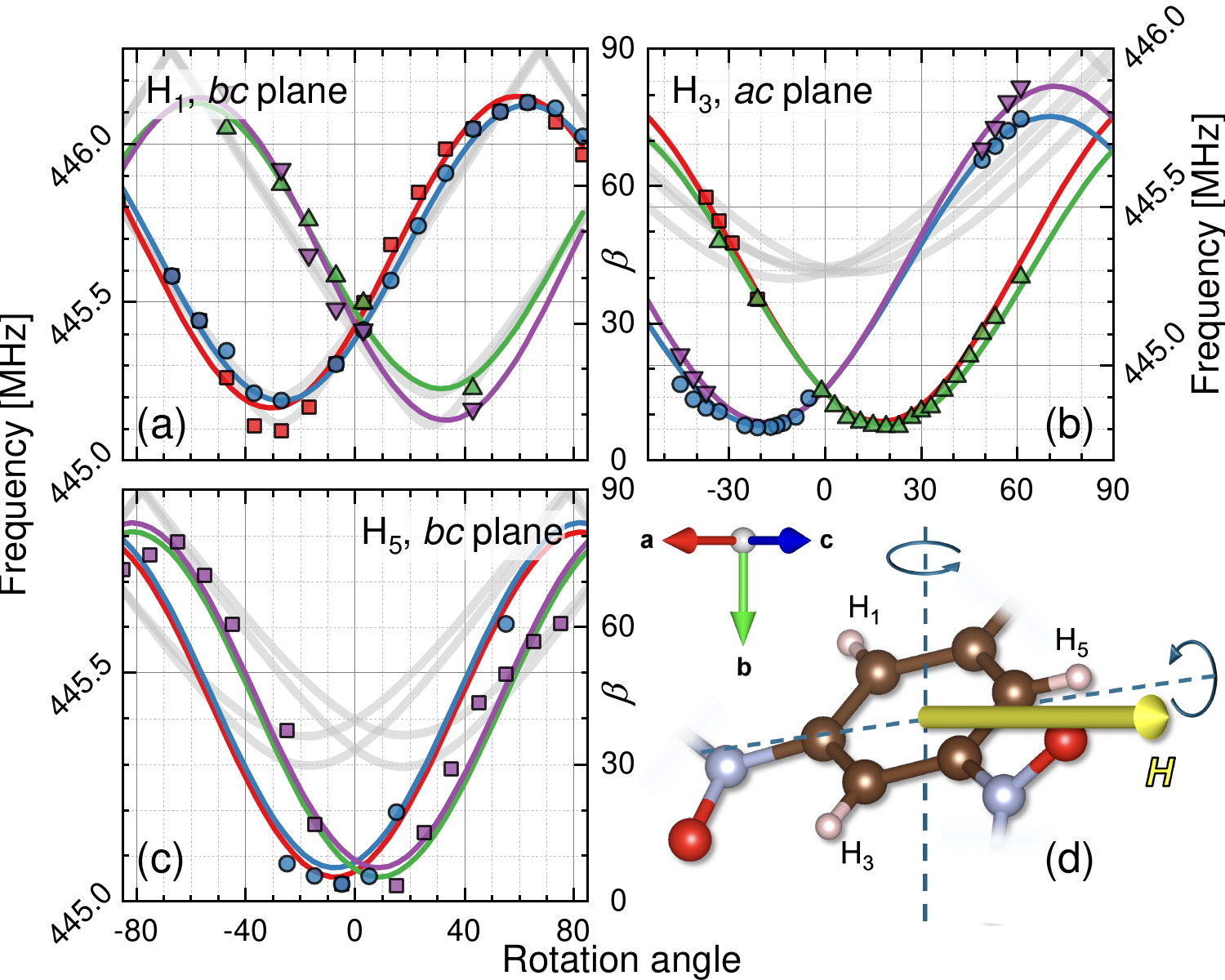}
	\caption{(a)-(c) Rotational positions of H$_1$, H$_3$, and H$_5$ NMR lines (separate panels used for clarity) in the paramagnetic phase at $T=20$~K, with $\mathbf{K}$ tensor fits for each site. When the magnetic field is rotated in the plane of symmetry ($ac$ or $bc$), only two lines are visible, however, \blue{a} small initial misalignment splits the spectra, revealing four lines for each site. Gray lines illustrate the rotational dependence of \blue{$\beta$}-  an angle between the C$_i$-H$_i$ bonds and the magnetic field \textbf{H}. (d) A schematic of the rotations is presented, together with the magnified view of the aromatic ring that connects the three hydrogen sites. There are 8 BoNO molecules in a unit cell, which gives four different orientations of the corresponding aromatic rings (Fig.~\ref{fig:fig2}). }
	\label{fig:rotation}
\end{figure}
\begin{equation}
	\mathbb{D} = \sum_i \frac{\mu_0g\mu_B}{4\pi |\textbf{r}_i|^3}\begin{pmatrix}
		3x_i^2-1 & 3x_i y_i & 3x_i z_i \\
		3x_i y_i & 3y_i^2 - 1 & 3y_i z_i \\
		3x_i z_i & 3y_i z_i & 3z_i^2 -1 
	\end{pmatrix},
\end{equation}
where $\textbf{r}_i = (x_i, y_i, z_i)$ is a position vector of a $S=1$ magnetic moment with respect to the $^1$H nucleus given in Cartesian coordinates. The sum goes over multiple unit cells until convergence is reached.  In \blue{the} case of BoNO, $\mathbb{D}$ is calculated by considering delocalized $S=1$ positions and weight-averaging their contributions using DFT calculated probability density function with a chosen convergence radius $r_c = 100$~\AA. Due to $1/r^3$ dependence of a dipolar coupling, convergence is reached rapidly within $r \approx 30$~\AA. The demagnetization effects $\mathbf{N}$ are usually neglected in paramagnetic compounds, however, in a highly-asymmetric rod-shaped crystals the contribution to the external field is of the order of the observed NMR shifts ($\approx 0.1 \%$). Similarly, the transferred hyperfine coupling should be negligible in a covalently bonded system with a strong electron localization, but we argue that \blue{in BoNO} the nature of delocalized orbitals induces a significant coupling between $^1$H site and the electron spins formally placed on NO molecular groups. The fitted values of the $\mathbf{K}$ tensor for H$_1$, H$_3$ and H$_5$ positions are summarized in Table~\ref{table}.  Indeed, as can be seen, the demagnetization and dipolar coupling alone cannot adequately explain the observed NMR frequency shifts and rotational dependence. By subtracting the dipolar and demagnetization components we show that the NMR shift is affected dominantly by the transferred hyperfine coupling mediated by the delocalized $\pi$-orbital of \blue{the} aromatic ring.\\
\indent We continue to the NMR spectrum site assignation. Typically, if the transferred hyperfine coupling were negligible, one could assign NMR lines to specific sites by comparing the behavior of the NMR shifts with rotational dependence of the  dipolar term. However, in our case\blue{,} we resort\blue{ed} to a different approach. It has been shown for a similar molecular environment~\cite{Southern,Orendt1997, Colhoun2002} that the largest hyperfine tensor component is dominantly oriented along (within $\pm 10^\circ$) the corresponding C$-$H bond, while the smallest component is oriented (again, within $\pm 10^\circ$) perpendicular to the aromatic ring. To assign each $\mathbf{K}$ tensor to the correct crystallographic position, we matched the observed rotation dependence of the shift to the rotation dependence  of the angle \blue{$\beta$} between each C$-$H bond and the external magnetic field (gray curves in Fig.~\ref{fig:rotation}~(a)-(c)), as depicted for a single molecule in Fig.~\ref{fig:rotation}~(d). Since the plane formed by H$_1$, H$_3$ and H$_5$ sites is tilted with respect to the rotation planes, for selected rotations\blue{,} the magnetic field will never be completely along the C$-$H bond, however,  as the sample rotates the angle \blue{$\beta$}will reach its minimum position (corresponding to the largest frequency shift) and approximately $90^\circ$ away from that position (the exact angle depends on the precise orientation of the C$-$H bond) the C$-$H bond will be perpendicular to the magnetic field. As shown, each rotational dependence of the shift consistently follows the rotational dependence of the angle \blue{$\beta$}, with almost matching minimal and maximal values. It should be kept in mind that in Fig.~\ref{fig:rotation}~(d) for clarity we show only one aromatic ring from the unit cell, for which one can observe one of the presented curves in (a)-(c).  \blue{The}  other three curves are created by other symmetry-related H$_1$, H$_3$ and H$_5$ sites in the unit cell.

\section{Conclusion}

In this work, we have presented a novel organic compound  \bono\ (abbreviated BoNO), as an ideal candidate for a prototype Haldane system. In BoNO, an $S=1$ \blue{unit is formed by} a strong ferromagnetic intramolecular coupling $|J_\text{FM}|\blue{/k_B} \gtrsim 500$~K between electrons from two radical NO molecular groups. We have determined that BoNO crystallizes in an orthorhombic \textit{Pbca} unit cell with a molecular stacking crucial for a quasi-one-dimensional behavior. \blue{The} bulk magnetization and EPR measurements in the paramagnetic phase display the Curie-Weiss behavior with the effective moment $\mu_\text{meas.} = (2.7 \pm 0.1)\mu_B$/f.u. corresponding to the $S=1$ system. Both measurements feature a maximum in susceptibility at $\approx 15$~K, followed by a steep suppression in $\chi(T)$ characteristic of low-dimensional systems. The paramagnetic $\chi(T)$ dependence suggest\blue{s} a dominant antiferromagnetic spin\blue{-}exchange which can be attributed to $J_\text{1D}\blue{/k_B} = (11.3\pm0.1)$~K obtained by fits to Padé approximated QMC data. At the high-temperature end, the fit to the \blue{biradical} susceptibility model is virtually indistinguishable from a Curie-Weiss law with $S=1$, proving that \blue{the}  ferromagnetic intramolecular coupling constant is much larger than 300~K. When the bulk magnetization measurements are performed at 1.3~K up to the $\mu_0H = 40$~T, we observe a quasi-linear increase above $\mu_0 H_\text{c1}\approx 2$~T where the singlet-triplet gap in the system is closed, and a fully polarized state above $\mu_0 H_\text{c2}\approx33$~T. These field values indicate that in this Haldane system the entire field-induced phase diagram can be thoroughly studied. The EPR measurements also confirm that BoNO has a highly isotropic $g$ tensor  in the paramagnetic state with $g_a = 2.0041$, $g_b = 2.0065$, and $g_c = 2.0063$ remarkably close to the free electron value $g_e = 2.0032$. At low temperatures\blue{,} $g_i$ ($i=a, b, c$) deviate from $g_e$  in accordance with emerging one-dimensional short\blue{-}range antiferromagnetic correlations. The EPR linewidth changes can be attributed to both dipolar and short-range order effects present in the sample. Finally, we did temperature\blue{-}dependent $^1$H NMR measurements to corroborate the previous findings. We have analyzed \blue{the} $^1$H NMR rotation spectra, assigned specific spectral lines to the crystallographic H$_1$, H$_3$ and H$_5$ positions and determined the magnitude and origin of the nuclear resonance frequency shifts. The presence of $^1$H sites which are strongly coupled to the $S=1$ spin system makes $^1$H NMR a valuable technique for the future study of the TLL and BEC physics in the \blue{system}. \blue{In particular, theoretical predictions for the TLL phase in a Haldane chain can now be tested in the entire phase space~\cite{Konik2002}. Furthermore, the study of emerging excitations~\cite{Mukhopadhyay2012} at the upper critical field $H_\text{c2}$ can be extended to a Haldane chain system.}  Demonstrably, BoNO presents a unique research opportunity as a Haldane system with no apparent spin Hamiltonian anisotropy and a textbook example of properties in a low-dimensional quantum spin system.

\section{Acknowledgments}
We acknowledge fruitful discussions with M. Dupont, N. Laflorencie, S. Capponi, N. Došlić and P. Sengupta. We thank D. Pajić and D. Barišić for helping us with the susceptibility measurements. M.S.G. and I.J. acknowledge the support of Croatian Science Foundation (HRZZ) under the project IP-2018-01-2970 and the support of project CeNIKS co-financed by the Croatian Government and the European Union through the European Regional Development Fund - Competitiveness and Cohesion Operational Programme (Grant No. KK.01.1.1.02.0013). This work was partly supported by JSPS KAKENHI Grant Number JP23K25824. M. Herak acknowledges the support of project  Cryogenic Centre at the Institute of Physics - KaCIF, co-financed by the Croatian Government and the European Union through the European Regional Development Fund – Competitiveness and Cohesion Operational Programme (Grant No. KK.01.1.1.02.0012) and the support of the project Ground states in competition – strong correlations, frustration and disorder — FrustKor financed by the Croatian Government and the European Union through the National Recovery and Resilience Plan 2021-2026 (NPOO).

\section*{APPENDIX A: Estimating intramolecular coupling $J_\text{FM}$}
\label{AppA}
\begin{figure}[!h]
	\includegraphics[width=\columnwidth]{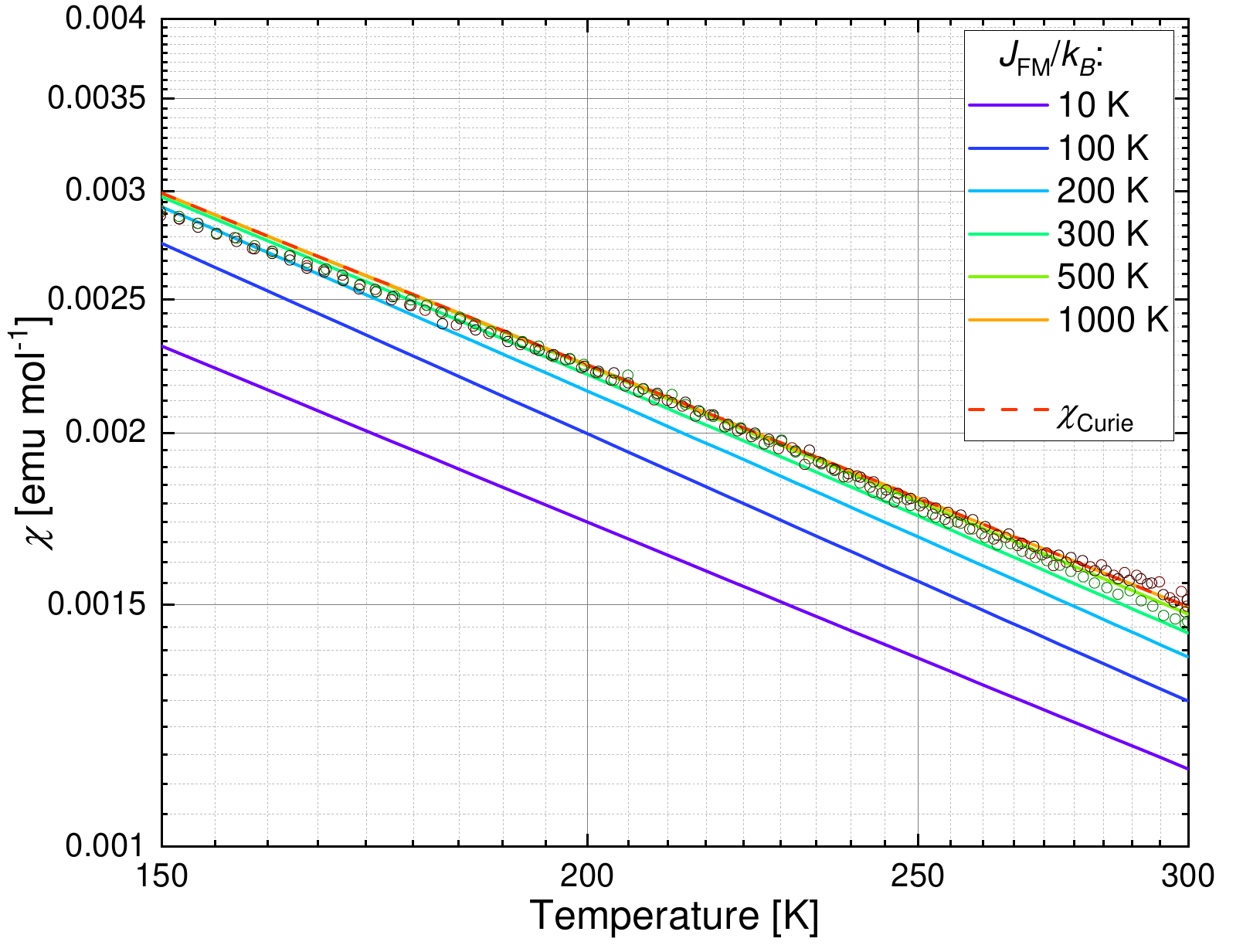}
	\caption{Comparison of susceptibility values (circles) at temperatures above 150~K to Eq.~(\ref{eq:Rajca}) for different $J_\text{FM}$ values and to the Curie temperature dependence from fits in Fig.~\ref{fig:fig3}. From the dataset one can conclude that $|J_\text{FM}|\blue{/k_B} \gtrsim 500$~K. }
	\label{fig:HTsucs}
\end{figure}
\blue{The} intramolecular coupling $J_\text{FM}$ in the organic \blue{biradicals} can be determined from the magnetic susceptibility measurements when $k_B T/|J_\text{FM}| \gtrsim 1 $~\cite{Rajca1994}:
\begin{equation}
	M_\text{\blue{birad}} = 2 g\mu_B \left(\frac{\sinh{\delta}}{1+2\cosh{\delta}+e^{-2|J_\text{FM}|/\blue{k_B}T}}\right),
	\label{eq:Rajca}
\end{equation}
where $\delta = g\mu_0\mu_BH/k_B(T-\Theta_\text{MF})$, and $\Theta_\text{MF}$ is a mean-field correction for the intermolecular interaction $J_\text{1D}$. This general expression reduces to the well-known Bleaney-Bowers equation~\cite{Bleaney1952} in low magnetic fields, $g\mu_0\mu_B H \ll k_B T$ or Curie-Weiss behavior in the strong coupling limit $|J_\text{FM}| \gg k_B T$. As is shown in Fig.~\ref{fig:HTsucs}, it was not  possible to distinguish between a Curie-Weiss and $M_\text{dirad}$ for $|J_\text{FM}|\blue{/k_B} \gtrsim 500$~K in our measurements, which leads to the conclusion that \blue{in BoNO} intramolecular coupling is much larger than 300~K. This corroborates our DFT calculation result that gave $|J_\text{FM}|\blue{/k_B} \approx 450$~K. In any case, below room temperature the system can be regarded as an $S=1$ chain.
\newpage 
	\bibliography{BibList}

\end{document}